\documentclass[journal]{IEEEtran}

\usepackage{amsmath,amssymb,amsfonts,amsthm,mathrsfs}
\usepackage{algorithm,algorithmic}
\usepackage{graphicx}
\usepackage{textcomp}
\usepackage{color,xcolor}
\usepackage{bm}
\usepackage{multirow}
\usepackage[normalem]{ulem}
\usepackage{tabularx}
\usepackage{xpatch}
\usepackage{soul}
\usepackage{ulem}
\usepackage{verbatim}
\usepackage{subcaption}
\usepackage{enumitem}
\usepackage{hyperref}
\usepackage[acronym]{glossaries}
\usepackage[linewidth=0.8pt]{mdframed}
\usepackage{etoolbox}
\usepackage{cite}

\hypersetup{
	colorlinks=true,  
	linkcolor=red,    
	citecolor=blue,   
	 urlcolor=black   
}

\theoremstyle{theorem,lemma,remark,proposition}

\newtheorem{remark}{Remark}

\ifCLASSINFOpdf
\else
\fi

\begin{document}

\title{
Polarization-Aware DoA Detection Relying on a Single Rydberg Atomic Receiver
}

\author{
Yuanbin~Chen,~Chau~Yuen,~\IEEEmembership{Fellow,~IEEE},~Darmindra~Arumugam,~Chong~Meng~Samson~See,~\IEEEmembership{Member,~IEEE},~M\'{e}rouane~Debbah,~\IEEEmembership{Fellow,~IEEE},~and~Lajos~Hanzo,~\IEEEmembership{Life~Fellow,~IEEE}


\thanks{
This work was supported in part by MoE AcRF Tier 1 Thematic Grant RT12/23 023780-00001. 
The portion of this work conducted at the Jet Propulsion Laboratory, California Institute of Technology, was carried out under a contract with the National Aeronautics and Space Administration (80NM0018D0004).
The financial support of the following Engineering and Physical Sciences Research Council (EPSRC) projects is gratefully acknowledged: Platform for Driving Ultimate Connectivity (TITAN) (EP/X04047X/1; EP/Y037243/1); Robust and Reliable Quantum Computing (RoaRQ, EP/W032635/1); PerCom (EP/X012301/1); India-UK Intelligent Spectrum Innovation ICON UKRI-1859.

Yuanbin Chen and Chau Yuen are with the School of Electrical and Electronics Engineering, Nanyang Technological University, Singapore 639798 (emails: yuanbin.chen@ntu.edu.sg; chau.yuen@ntu.edu.sg).

Darmindra~Arumugam is with the Jet Propulsion Laboratory, California Institute of Technology, Pasadena, 91109, California, USA (email: darmindra.d.arumugam@jpl.nasa.gov).

Chong Meng Samson See is with DSO National Laboratories, Singapore 118225 (e-mail: schongme@dso.org.sg).

M\'{e}rouane Debbah is with the Research Institute for Digital Future, Khalifa University, 127788 Abu Dhabi, UAE (e-mail: merouane.debbah@ku.ac.ae).

Lajos Hanzo is with School of Electronics and Computer Science, University of Southampton, SO17 1BJ Southampton, U.K. (e-mail: lh@ecs.soton.ac.uk).	

}
\vspace{-1em}
}

\maketitle

\begin{abstract}

A polarization-aware direction-of-arrival~(DoA) detection scheme is conceived that leverages the intrinsic vector sensitivity of a single Rydberg atomic vapor cell to achieve quantum-enhanced angle resolution. Our core idea lies in the fact that the vector nature of an electromagnetic wave is uniquely determined by its orthogonal electric and magnetic field components, both of which can be retrieved by a single Rydberg atomic receiver  via electromagnetically induced transparency~(EIT)-based spectroscopy. To be specific, in the presence of a static magnetic bias field that defines a stable quantization axis, a pair of sequential EIT measurements is carried out in the same vapor cell.
Firstly, the electric-field polarization angle is extracted from the Zeeman-resolved EIT spectrum associated with an electric-dipole transition driven by the radio frequency~(RF) field. Within the same experimental cycle, the RF field is then retuned to a magnetic-dipole resonance, producing Zeeman-resolved EIT peaks for decoding the RF magnetic-field orientation. This scheme exhibits a dual yet independent sensitivity on both angles, allowing for precise DoA reconstruction without the need for spatial diversity or phase referencing. Building on this foundation, we derive the quantum Fisher-information matrix (QFIM) and obtain a closed-form quantum Cram\'{e}r-Rao bound (QCRB) for the joint estimation of polarization and orientation angles. Finally, simulation results spanning various quantum parameters validate the proposed approach and identify optimal operating regimes. With appropriately chosen polarization and magnetic-field geometries, a single vapor cell is expected to achieve sub-0.1$^\circ$ angle resolution at moderate RF-field driving strengths.

\end{abstract}


%
\IEEEpeerreviewmaketitle

\section{Introduction}

The emerging blueprint for sixth-generation (6G) communications calls for centimeter-level positioning accuracy, sub-millisecond latency and pervasive sensing at millimeter-wave and Terahertz frequencies - a performance that pushes today's electromagnetic (EM) receiver technology to its classical limits. Against this backdrop, quantum sensing emerges like dawn-light piercing through the fog of the physical layer designs~\cite{QS-101}. In particular, Rydberg atomic receivers exploit the extreme sensitivity of highly-excited alkali atoms to convert incident radio-frequency (RF) fields directly into optical signatures, enabling simultaneous resolution of electric-field amplitude, phase, and polarization information at sub-wavelength scales~\cite{terry-WCM,QS-22,QS-3,QS-31,Yuan_2023}.
This capability of exceeding classical detection theory limits  redefines the paradigm of joint multi-dimensional EM parameter extraction.

Owing to their large electric-dipole moments, highly excited Rydberg atoms lend themselves to coupling resonantly with microwave and Terahertz radiation, allowing them to respond to electric fields across a broad frequency spectrum having extreme sensitivity. The high sensitivity of single-cell Rydberg atomic receivers has bee experimentally shown to be on the order of $\mu$V/cm/$\sqrt{\text{Hz}}$ using a standard structure~\cite{QS-3,QS-12,QS-15}. This has also been improved to $55~\text{nV/cm/}\sqrt{\text{Hz}}$ based on a superheterodyne structure~\cite{QS-16}, and the employment of the six-wave mixing technique further boosts its sensitivity to $3.98~\text{nV/cm/}\sqrt{\text{Hz}}$~\cite{QS-30}. The sensitivity limit of single-cell Rydberg atomic receivers may even reach $\text{pV/cm/}\sqrt{\text{Hz}}$ under standard quantum limit conditions~\cite{QS-31}.
Optical readout of the Rydberg electromagnetically induced transparency (EIT) or Autler-Townes (AT) response allows simultaneous  extraction of the incident field amplitude~\cite{QS-12,QS-15,QS-20,QS-9-TAP-2024,QS-1}, of polarization information~\cite{QS-24,QS-26,QS-54,QS-51}, as well as of both the magnitude and orientation of a magnetic field~\cite{QS-52,QS-53}. Introducing an RF local oscillator (LO) further eases the direct recovery of the signal phase~﻿\cite{QS-16,QS-27,QS-28,chen-arxiv,terry-trans}. We now provide a concise overview of these operating principles and state-of-the-art advances in the field of wireless sensing.

\textit{Electric-Field Amplitude Measurement:}
The EIT technique is typically employed for microwave metrology. When the RF field applied is resonant with the transition frequency between two Rydberg states, the EIT signal exhibits a characteristic splitting phenomenon known as AT splitting~\cite{QS-22}. The strength of the electric field can be directly extracted from the AT splitting interval in the EIT spectrum. Upon introducing an LO, the atomic system acts as a Rydberg atomic mixer, enabling the EIT/AT effects in Rydberg atoms to demodulate a second, co-polarized RF field. The resultant difference frequency, or intermediate frequency, can be optically detected through optically probing the Rydberg atoms~﻿\cite{QS-16,QS-27,QS-28,chen-arxiv,terry-trans}.

\textit{Electric-Field Polarization Measurement:} The EIT spectroscopy also supports the polarization measurement of the electric field. The technique presented in~\cite{QS-24} employs Rydberg-atom based EIT in rubidium vapor cells to measure the electric-field vector, reaching an angle resolution of $0.5^\circ$. In~\cite{QS-26}, the employment of an additional LO facilitates the construction of a Rydberg-atom based mixer, through which a polarization resolution below $0.5^\circ$ can be attained. 
Furthermore, it is established in~\cite{QS-51} that for a degenerated $S_{1/2} \to P_{1/2}$ manifold driven by a linearly polarized RF field, the AT splitting and underlying eigen-energies are invariant when the polarization axis is rotated, which is actually an immediate consequence of the spherical symmetry of the Hamiltonian. In \cite{QS-54}, this symmetry can be deliberately broken, e.g., by applying a weak bias field, in which case the AT line-shape becomes polarization-dependent. Such mild symmetry breaking inspires the use of variations in $\pi$ and $\sigma^\pm$ coupling strengths for RF vector sensing.

\textit{Magnetic-Field Measurement:} The measurement of magnetic fields, including both their magnitude and orientation, can also be pursued under the same framework. Specifically, EIT resonances are associated with the preparation of a coherent, non-interacting superposition state (a so-called ``dark state"), which typically involves several metastable atomic levels, such as the Zeeman sub-levels or hyperfine sub-levels of alkali atoms. In the presence of a pair of probe and coupling optical fields, a twin-photon Raman resonance in a $\Lambda$ configuration is established, where the magnetic field strength can thus be inferred from Zeeman shifts within the EIT resonance~\cite{QS-55-2003}. In~\cite{QS-58}, magnetic-field-induced splitting of the EIT resonance is observed in a three-level ladder system, modeling the vector response to an RF-modulated magnetic field. Furthermore, the orientation of the static bias-field can be extracted by analyzing the relative strengths of the EIT peaks \cite{QS-52, QS-53}.
More specifically, it is demonstrated in \cite{QS-52, QS-53} that in a $\text{lin} \parallel \text{lin}$ optical EIT vector-magnetometry configuration, the EIT resonance amplitude depends on the particular orientation of a static bias magnetic field. Here, we denote the laser wave vector by $\mathbf{k}_\text{opt}$ and the laser polarization by ${\bm{\epsilon}}_\text{L}$. The resonance amplitude reaches an extremum (typically a maximum) when ${\bm{\epsilon}}_\text{L} \perp \text{span} \left\lbrace {{\mathbf{B}}_\text{bias},  \mathbf{k}_\text{opt} }\right\rbrace  $, equivalently, $ {\bm{\epsilon}}_\text{L} \parallel \left(  \mathbf{k}_\text{opt} \times {\mathbf{B}}_\text{bias}   \right) $. Therefore, by measuring the EIT amplitude while rotating the laser polarization around a fixed $\mathbf{k}_\text{opt}$, one can retrieve the orientation of $\mathbf{B}_\text{bias}$ (the EIT remains observable without any RF field; only its contrast varies with geometry).


\textit{Direction-of-Arrival (DoA) Detection:} Several studies have leveraged spatially distributed Rydberg atomic receivers to detect the DoA or angle of arrival (AoA) of incident RF fields~\cite{QS-11, QS-41, QS-56, terry-wcl,Yan_2023}. Despite differences in implementation, the core idea in~\cite{QS-11, QS-41, QS-56} is fundamentally similar to classical antenna arrays: they utilize multiple vapor cells~\cite{QS-11}, or arrange multiple probe and coupling beams into laser arrays~\cite{QS-41,terry-wcl}, in order to receive RF signals at various spatial positions. By analyzing the amplitude and/or phase differences of the received signals, these treatises apply direction-finding algorithms for retrieving the DoA of the RF source.

Despite these impressive advances, existing Rydberg-based DoA estimation schemes still face critical limitations. Firstly, their angle resolution inherently scales with the number of sensing elements, but increasing that number is non-trivial. Deploying multiple vapor cells~\cite{QS-1,QS-41} requires phase-coherent detection and meticulous cross-calibration, while each cell's response is susceptible to local temperature and pressure drifts. Regarding the laser-array based approach implemented in a single vapor cell~\cite{terry-wcl,QS-9-TAP-2025}, accurate beam alignment is a requisite and optical cross-talk may significantly erode the performance. Secondly, phase-sensitive techniques rely on an additional LO field; nevertheless, reflections from scatterers may distort the phase-difference-to-DoA mappings. As a remedy, the study in~\cite{QS-56} investigates a metal-plate-integrated vapor cell using only a probe-coupling laser pair, achieving 
an angle resolution  of $1.7^\circ$ without the need for a secondary RF phase reference field.

Inspired by the Rydberg atomic receiver's intrinsic capability of detecting both polarization and magnetic-field orientation, we propose a DoA detection paradigm using only a probe-coupling laser pair in a single vapor cell. Our approach exploits a fundamental property of plane electromagnetic waves, that is, the wave propagation direction $\mathbf{k}$ is uniquely determined by the orthogonal electric $\mathbf{E}$ and magnetic $\mathbf{B}_{\text{RF}}$ field vectors. By correlating the Zeeman-resolved EIT spectral responses with both the $\mathbf{E}$-polarization (via electric-dipole transition) and $\mathbf{B}_{\text{RF}}$-orientation (via magnetic-dipole transition), the DoA can be reconstructed without resorting to spatial diversity or phase referencing. This design harnesses the Rydberg atoms' native vectorial sensitivity, circumventing the scalability bottlenecks of conventional quantum sensor arrays. The key contributions of this work are summarized as follows.

\begin{itemize}
\item 

We develop an analytical technique for determining the DoA by decoding both the electric-field polarization and Zeeman-resolved magnetic sub-level structure using EIT spectroscopy. A pair of back-to-back EIT measurements is performed in the same single vapor cell and referenced to a single, static bias field that defines the quantization axis for the experimental cycle. 
Firstly, given an electric-dipole (E1) transition driven by the RF field, the polarization angle $\theta_{{\text{RF}}}$ can be extracted from the Zeeman-resolved EIT spectrum. Within the same experimental cycle, the carrier frequency is retuned for driving a magnetic-dipole (M1) transition. The resultant Zeeman-resolved EIT peaks are determined by the spherical-basis components of the RF magnetic field with respect to the quantization axis; the spectra taken for two bias-field orientations yield the RF magnetic-field orientation $\theta_{B}$. The proposed scheme exhibits a dual yet independent relationship with this angle pair, facilitating precise DoA reconstruction by jointly addressing these angular dependencies.

\item We place the angle-resolution capability of a Rydberg atomic receiver on the same theoretical footing as that of classical antenna arrays. Specifically, commencing from the quantum Fisher-information matrix (QFIM), we derive the quantum Cram\'{e}r-Rao bound (QCRB) for the joint estimation of the electric-field polarization angle $\theta_{\text{RF}}$ and the magnetic-field orientation $\theta_B$. Since the DoA detection accuracy of our scheme is ultimately limited by how precisely one can read the EIT spectrum, we identify the squared gradient of the probe-transmission profile as the relevant signal term and define an information-centric signal-to-noise ratio (SNR). This thus leads, under weak-probe conditions, to a compact closed-form QCRB that depends only on the SNR, enabling a direct comparison with the Cram\'{e}r-Rao bound (CRB) of a conventional uniform linear array (ULA).

\item We conduct our quantum-domain simulations using the open-source QuTip quantum-optics toolkit~\cite{QuTip}, in order to validate the proposed analytical DoA-detection scheme across a number of quantum parameters. We find that, for appropriately chosen polarization and magnetic geometries, a single vapor cell delivers sub-$0.1^{\circ}$ angle resolution at moderate RF-field driving strengths. However, we also show that either over-driving or under-driving the RF transition erodes performance. When benchmarked against array-based schemes, such as ULA~\cite{ULA-CRB1} and vector sensor array~\cite{VSA}, a single Rydberg atomic receiver outperforms a 16-element array across a low-SNR regime. The QCRB saturates at the SQL floor, beyond which the angle resolution no longer improves upon increasing the SNR, while further enhancement in detection precision can by achieved via tuning intrinsic quantum parameters of the system.

\end{itemize}

The remainder of this paper is structured as follows. Section~\ref{Sec_II} presents the core idea of our proposed scheme. Section~\ref{Sec_III} and Section~\ref{Sec_IV} demonstrate how the Zeeman-resolved EIT spectrum can be processed for extracting the polarization angle and RF magnetic-field orientation, respectively. Next, Section~\ref{Sec_V} quantifies the quantum resolution limits via the QCRB performance. Our simulation results are provided in Section~\ref{Sec_VI}, and finally, we conclude in Section~\ref{Sec_VII}.

\textit{Notation:} Throughout this paper, we adopt the following mathematical conventions. In particular, ${\text{Tr}}\left(  \cdot  \right)$ denotes the trace of a matrix; ${\left(  \cdot  \right)^*}$ and ${\left(  \cdot  \right)^H}$ represent the complex conjugate and conjugate transpose (Hermitian transpose) of a scalar, vector, or matrix, respectively; ${\left(  \cdot  \right)^{ - 1}}$ denotes the matrix inverse; ${\text{vec}}\left(  \cdot  \right)$ is the vectorization operator that stacks the columns of a matrix into a column vector; ${\left\|  \cdot  \right\|_F}$ signifies the Frobenius norm of a matrix. The symbol $\propto$ denotes the proportionality; $\otimes$ and $\oplus$ represents the Kronecker product and the direct sum between two matrices. Regarding quantum‐mechanical notations, Dirac's bra-ket formalism represents quantum states by kets $\left| \psi  \right\rangle $ and bras $\left\langle \varphi  \right|$. The expression $\left\langle {\varphi \left| M \right|\psi } \right\rangle $ denotes the matrix element of an operator between the initial state $\left| \psi  \right\rangle $ and the final state $\left\langle \varphi  \right|$. Furthermore, a reduced matrix element is denoted by double bars, e.g., $\left\langle { m^\prime_J\left\| T \right\|{m_J}} \right\rangle  $, where the double‐bar notation indicates independence from the magnetic quantum numbers.

\section{Core Idea}\label{Sec_II}
The core idea in determining the DoA of the incoming RF signal using a single Rydberg atomic receiver relies on the fact that a plane electromagnetic wave is fully specified by a pair of mutually-orthogonal vectors, i.e., the electric-field vector $\mathbf{E}$ and the oscillating RF magnetic-field vector $\mathbf{B}_{\text{RF}}$. By resorting to the electric-field vector $\mathbf{E}$ and the oscillating RF magnetic-field vector $\mathbf{B}_{\text{RF}}$, the propagation vector $\mathbf{k}$ can be uniquely determined, specifically, ${\mathbf{k}} = \frac{{{\mathbf{E}} \times {\mathbf{B}_{\text{RF}}}}}{{\left\| {{\mathbf{E}} \times {\mathbf{B}_\text{RF}}} \right\|}}${\footnote{This reconstruction assumes the presence of a dominant radiative plane-wave component, so that the local field is approximately transverse. In strong near-field or comparable-power multipath conditions, the estimate should be interpreted as the local energy-flow direction, which is beyond the single-point and plane-wave-based formulation studied here.}}. The Rydberg atomic receiver delivers these two angles through two independent spectroscopic signatures recorded in the presence of a known static magnetic bias field $\mathbf{B}_{\text{bias}}$ that defines the quantization axis and delivers Zeeman-resolved magnetic sub-levels. 

\begin{itemize}
	\item \textit{Electric-field polarization angle} $\theta_{\text{RF}}$: We determine $\theta_{\text{RF}}$ via the electric-dipole $L=1$ (E1) transition $n^\prime S_{1/2} \to nP_{1/2}$, which is addressed by the RF field in presence of a bias field $\mathbf{B}_{\text{bias}}$. 
	The static magnetic bias field $\mathbf{B}_{\text{bias}}$ lifts the degeneracy of magnetic sub-levels via the Zeeman effect~\cite{Zeeman_shift}, producing a resolvable multi-peak structure in the EIT spectrum.
	As a result, the excited Rydberg states $n^\prime S_{1/2}$ and $nP_{1/2}$ are split into Zeeman sub-levels $\left\{ {\left| {{n^\prime }{S_{1/2}},m_J^\prime } \right\rangle ,\left| {n{P_{1/2}},{m_J}} \right\rangle } \right\}$, which can be characterized by the Rabi frequency of the E1-transition $\Omega_{{\text{E1}}}$, scaling with the projection of $\mathbf{E}$ onto the atomic transition dipole~$\hat {\mathbf{d}}$
	\begin{equation}\label{Omega_RF_reduce}
{\Omega _{{\text{E1}}}}\left( {{\theta _{{\text{RF}}}}} \right) \propto \frac{{\left| { \hat{\mathbf{d}} \cdot {\mathbf{E}}\left( {{\theta _{{\text{RF}}}}} \right)} \right|}}{\hbar },
	\end{equation}
	where $\hbar$ is the reduced Planck's constant. By measuring the relative AT splittings associated with linearly ($\pi$)-polarized and circularly ($\sigma^\pm$)-polarized components, we directly extract the electric-field polarization angle $\theta_{\text{RF}}$.
	
	\item \textit{Oscillating RF magnetic-field orientation} $\theta_B$: The orientation $\theta_B$ can be determined via the magnetic-dipole $L=1$ (M1) transition $n^\prime P_{1/2} \to nP_{3/2}$ driven by an RF electric field in the presence of a bias field $\mathbf{B}_{\text{bias}}$. Similarly,  the bias field facilitates both excited Rydberg states to be split into Zeeman sub-levels $\left\{ {\left| {{n^\prime }{P_{1/2}},m_J^\prime } \right\rangle ,\left| {n{P_{3/2}},{m_J}} \right\rangle } \right\}$, and defines $\pi$-polarized and $\sigma^\pm$-polarized selection rules. Then, the amplitudes of the Zeeman-resolved M1 peaks, which are characterized by the Rabi frequency of the M1-transition ${\Omega _{{\text{M1}}}}\left( { \theta_B } \right)$, scale as
	\begin{equation}
{\Omega _{{\text{M1}}}}\left( {{\theta _B}} \right) \propto \frac{{\left| {\hat{\bm{\mu}} \cdot {{\mathbf{B}}_{{\text{RF}}}}} \right|}}{\hbar },
	\end{equation}
	where $\hat{\bm{\mu}}$ is the magnetic dipole.	The oscillating RF magnetic-field  orientation $\theta_B$ can be identified when $\theta_{\text{bias}}$ is varied.
	
\end{itemize}

Given that we have $\mathbf{E} \perp \mathbf{B}_{\text{RF}}$ for a plane wave in the free space, knowing the angle pair $\left(\theta_{\text{RF}}, \theta_B \right) $ is sufficient for reconstructing the full three-dimensional DoA. 
Once these two vectors are identified, the propagation vector $\mathbf{k}$ follows from ${\mathbf{k}} = \frac{{{\mathbf{E}} \times {{\mathbf{B}}_{{\text{RF}}}}}}{{\left\| {{\mathbf{E}} \times {{\mathbf{B}}_{{\text{RF}}}}} \right\|}}$.
A single Rydberg atomic receiver can provide both vectors through a pair of independent spectroscopic signatures observed under the same static bias field $\mathbf{B}_{\mathrm{bias}}$.
The remainder of this paper derives analytically how $\theta_{\text{RF}}$ and $\theta_B$ imprint themselves on the EIT spectrum of the atomic system considered. 

\begin{figure*}[t]
	\centering
	\includegraphics[width=\textwidth]{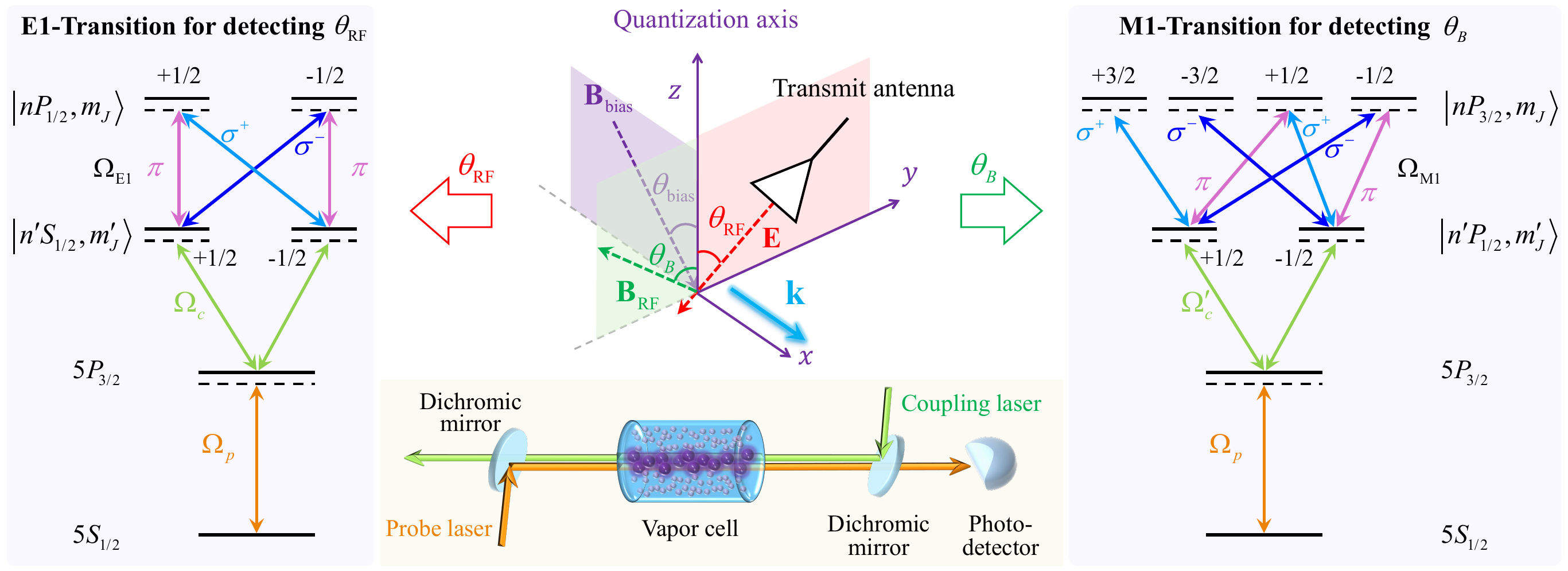}
	\caption{Illustration of the atomic system in the quantization axis specified by a static magnetic bias field $\mathbf{B}_{\text{bias}}$. The electric-field polarization angle $\theta_{{\text{RF}}}$ is inferred from the E1-transition and the RF magnetic-field orientation $\theta_{B}$ is obtained via the M1-transition.} \label{system_model}
\end{figure*}

\section{Detection of $\theta_{\text{RF}}$}\label{Sec_III}

In this section, we present a spectroscopic methodology for retrieving the electric‐field polarization angle $\theta_\text{RF}$ within a single Rydberg-atom vapor cell. 
We commence by modeling a four-level atomic system and derive its Hamiltonian. Next, we construct pure $\Lambda$-type configurations to resolve the Zeeman sub-levels~\cite{QS-53}, and derive the analytical expression for an E1-transition's Rabi frequency. Then, based on this, the polarization angle $\theta_\text{RF}$ can be extracted from the Zeeman-resolved peaks.

\subsection{Four-Level Atomic System and Hamiltonian}
We consider a four-level atomic system formulated as
\begin{equation}\label{E1_transition}
5{S_{1/2}}\mathop  {\longrightarrow} \limits^{{\Omega _p}} 5{P_{3/2}}\mathop  {\longrightarrow} \limits^{{\Omega _c}} n'{S_{1/2}}\mathop  {\longrightarrow} \limits^{{\Omega _{{\text{E1}}}}} n{P_{1/2}},
\end{equation}
where a probe laser (Rabi frequency $\Omega_p$) couples the ground state $5{S_{1/2}}$ to the intermediate excited state $5{P_{3/2}}$, and a strong coupling laser $\Omega_c$ drives the transition from $5{P_{3/2}}$ up to a Rydberg state $n'{S_{1/2}}$. Then, the RF field addresses the E1-transition $n'{S_{1/2}} \to n{P_{1/2}}$ whose Rabi frequency is denoted by $\Omega_{{\text{E1}}}$.  In this configuration, the probe transmission opens an EIT window whose AT splitting and Zeeman shifts provide direct spectroscopic access to both the amplitude/polarization of the RF field and the orientation of the oscillating RF magnetic field. As illustrated in Fig.~\ref{system_model}, we adopt a right-hand Cartesian laboratory frame in which the azimuth angle $0^\circ$ lies along the $y$-axis. In this configuration, the electric-field vector $\mathbf{E}$ lies in the $y-z$ plane and subtends an angle $\theta_{\text{RF}}$ with respect to the $z$-axis, while a static magnetic bias field $\mathbf{B}_{\text{bias}}$ defines a known angle $\theta_{\text{bias}}$ to the same reference. Note that for the measurement of the polarization angle $\theta_{{\text{RF}}}$, a bias field $\mathbf{B}_{\text{bias}}$ is not strictly required, since the probe-coupling beam geometry can itself define the quantization axis. However, we still retain a weak static bias field $\mathbf{B}_{\text{bias}}$ in the $x-z$ plane (yielding an angle $\theta_\text{bias}$ with the positive $z$-axis) such that the same well-defined quantization axis is used consistently in all electric- and magnetic-field measurements.
In our configuration, both the RF wave-vector and the co-propagating probe-coupling beams are aligned along the positive $x$-axis, which is orthogonal to the plane that contains both $\mathbf{E}$ and $\mathbf{B}_\text{RF}$. 


The coupling term between the RF field and the atomic dipole moment $\hat{\mathbf{d}}$ can be expressed as the Hamiltonian ${H_{{\text{int}}}}$, which is given by
\begin{equation}\label{H_int}
{H_{{\text{int}}}} =  - \hat{\mathbf{d}} \cdot {\mathbf{E}} = - \left\langle {n{P_{1/2}},{m_J}\left| {\hat{\mathbf{d}} \cdot {\mathbf{E}}} \right|{n^\prime }{S_{1/2}},{m_J^\prime}} \right\rangle  .
\end{equation}
Under the rotating wave approximation (RWA)~\cite{QS-51,QS-54}, the Hamiltonian of the four-level system considered can be formulated as
\begin{equation}
\resizebox{\hsize}{!}{$
{H_{{\text{int}}}} = \frac{\hbar }{2}\left( {\begin{array}{*{20}{c}}
		0&{ - {\Omega _p}}&0&0 \\ 
		{ - \Omega _p^*}&{ - {\Delta _p}}&{ - {\Omega _c}}&0 \\ 
		0&{ - \Omega _c^*}&{ - \left( {{\Delta _p} + {\Delta _c}} \right)}&{ - {\Omega _{{\text{RF}}}}} \\ 
		0&0&{ - \Omega _{{\text{RF}}}^*}&{ - \left( {{\Delta _p} + {\Delta _c} + {\Delta _{{\text{RF}}}}} \right)} 
\end{array}} \right),
$}
\end{equation}
where $\Delta_{p,c,\text{RF}}$ represents the detuning associated with the probe laser, coupling laser, and RF field, respectively. Since the RF field generally contains both a component parallel to the quantization axis and components that rotate around that axis, the electric field can be decomposed into the standard spherical basis $\left( {\bm{\epsilon}}_{-1}, {\bm{\epsilon}}_{0}, {\bm{\epsilon}}_{+1}\right) $ defined with respect to the quantization axis set by the bias field $\mathbf{B}_\text{bias}$. Here ${\bm{\epsilon}}_{0}$ represents the $\pi$-polarized unit vector that drives $\Delta m = 0$ transitions, whereas $\sigma^{\pm}$ correspond to the $\sigma^{\pm}$-circularly-polarized unit vectors that drive $\Delta m = \pm 1$ transitions. When expressing the spherical basis $\left( {\bm{\epsilon}}_{-1}, {\bm{\epsilon}}_{0}, {\bm{\epsilon}}_{+1}\right) $ in the right-handed Cartesian coordinate system illustrated in Fig.~\ref{system_model}, we obtain
${\bm{\epsilon}}_{0} = + \mathbf{x},~  {\bm{\epsilon}}_{+1} = \mp \frac{\mathbf{y} \pm \jmath \mathbf{z}}{\sqrt{2}}$,
where $\mathbf{x}$, $\mathbf{y}$, and $\mathbf{z}$ represent the unit vectors along the $x$, $y$, and $z$ axes, respectively.
The RF electric field can therefore be restructured as 
\begin{equation}
	{\mathbf{E}} = {E_0}\left[ {{\alpha _{ - 1}}\left( {{\theta _{{\text{RF}}}}} \right){ {\bm{\epsilon}} _{ - 1}} + {\alpha _0}\left( {{\theta _{{\text{RF}}}}} \right){  {\bm{\epsilon}}_0} + {\alpha _{ + 1}}\left( {{\theta _{{\text{RF}}}}} \right){{\bm{\epsilon}}_{ + 1}}} \right],
\end{equation}
where $E_0$ is the electric-field amplitude and the dimensionless coefficients ${\alpha _q} \left( q = 0, \pm 1\right) $ obey $\sum\nolimits_{q =  - 1}^{+1} {{{\left| {{\alpha _q}} \right|}^2} = 1} $. If we focus on the analysis of the $y-z$ plane, i.e., the plane orthogonal to the wave-vector $\mathbf{k} \parallel \mathbf{x}$, the projection of the field onto the spherical basis is purely determined by the angle $\theta_{{\text{RF}}}$ between $\mathbf{E}$ and the $y$-axis
\begin{align}
{\alpha _0}{\mkern 1mu} \left( {{\theta _{{\text{RF}}}}} \right) = \cos {\theta _{{\text{RF}}}}, \ {\alpha _{ \pm 1}}{\mkern 1mu} \left( {{\theta _{{\text{RF}}}}} \right) =  \mp \frac{{\sin {\theta _{{\text{RF}}}}}}{{\sqrt 2 }}.
\end{align}
These definitions explicitly indicate which components of the RF field couple to $\pi$ ($\Delta m = 0$) or to $\sigma^\pm$ ($\Delta m = \pm 1$) transitions, clarifying the role of $\theta_{{\text{RF}}}$ as the polarization angle of interest.

Referring to (\ref{H_int}) again, the dipole $\hat{\mathbf{d}}$ can be represented as a linear combination of 
rank-1 spherical tensor components ${\hat d_q}\left( {q = 0, \pm 1} \right)$, which is given by
\begin{equation}
	\hat {\mathbf{d}} = \sum\limits_{q =  - 1}^{ + 1} {{{\hat d}_q}{{\bm{\epsilon}}_q}} .
\end{equation}
The transition matrix elements in (\ref{H_int}) can be further reformulated as
\begin{align}\label{H_int_2}
 &- \left\langle {n{P_{1/2}},{m_J}\left| {\hat{\mathbf{d}} \cdot {\mathbf{E}}} \right|{n^\prime }{S_{1/2}},{m_J^\prime}} \right\rangle  
 \nonumber\\
  = &  - {E_0}\sum\limits_{q =  - 1}^{ + 1} {{\alpha _q}\left( {{\theta _{{\text{RF}}}}} \right)\left\langle {n{P_{1/2}},{m_J}\left| {{{\hat d}_q}} \right|{n^\prime }{S_{1/2}},{m_J^\prime}} \right\rangle } .
\end{align}
In (\ref{H_int_2}), $\left\langle {n{P_{1/2}},{m_J}\left| {{{\hat d}_q}} \right|{n^\prime }{S_{1/2}},m_J^\prime } \right\rangle $ can be expanded according to the Wigner-Eckart theorem as follows~\cite{W-E_theorem}
\begin{align}
&\left\langle {n{P_{1/2}},{m_J}\left| {{{\hat d}_q}} \right|{n^\prime }{S_{1/2}},m_J^\prime } \right\rangle \nonumber\\
 &= \left\langle {1/2,{m_J};1,q|1/2,m_J^\prime } \right\rangle \left\langle {n{P_{1/2}}\left\| {\hat d_q} \right\|{n^\prime }{S_{1/2}}} \right\rangle \nonumber\\
 &
 \resizebox{\hsize}{!}{$
 = \sqrt 2  \times {\left( { - 1} \right)^{1/2 - {m_J}}}\left( {\begin{array}{*{20}{c}}
 		{1/2}&1&{1/2} \\ 
 		{ - {m_J}}&q&{m_J^\prime } 
 \end{array}} \right)\left\langle {n{P_{1/2}}\left\| {\hat d_q} \right\|{n^\prime }{S_{1/2}}} \right\rangle ,
$}
\end{align}
where $\left\langle {n{P_{1/2}}\left\| {\hat d_q} \right\|{n^\prime }{S_{1/2}}} \right\rangle $ is the reduced dipole matrix element, which is dependent on the magnetic quantum number $m_J, m_J^\prime$, and the spherical component index $q$. The term $\left\langle {1/2,{m_J};1,q|1/2,{m_J}^\prime } \right\rangle $ is the Clebsch-Gordan coefficient~\cite{C-G_coefficients}, which can be  expanded in the form of a Wigner $3-j$ symbol formulated as~\cite{3-j}
\begin{align}\label{C-G_coefficient}
&\left\langle {1/2,{m_J};1,q|1/2,m_J^\prime } \right\rangle \nonumber\\
&  = \sqrt 2  \times {\left( { - 1} \right)^{1/2 - {m_J}}}\left( {\begin{array}{*{20}{c}}
		{1/2}&1&{1/2} \\ 
		{ - {m_J}}&q&{m_J^\prime } 
\end{array}} \right).
\end{align}
The Clebsch-Gordan coefficients in (\ref{C-G_coefficient}) characterize which transition $\Delta m$ is allowed and their own contributions. Consequently, (\ref{H_int_2}) can be explicitly written as
\begin{align}
 &- \left\langle {n{P_{1/2}},{m_J}\left| {\hat{\mathbf{d}} \cdot {{\mathbf{E}}}} \right|{n^\prime }{S_{1/2}},m^\prime_J } \right\rangle \nonumber\\
  &=  - {E_0}\sum\limits_{q =  - 1}^{ + 1} {{\alpha _q}\left( {{\theta _{{\text{RF}}}}} \right)\left\langle {n{P_{1/2}},{m_J}\left| {{{\hat d}_q}} \right|{n^\prime }{S_{1/2}},m^\prime_J } \right\rangle } \nonumber\\
  & =  - {E_0}\sum\limits_{q =  - 1}^{ + 1} {{\alpha _q}\left( {{\theta _{{\text{RF}}}}} \right){{\left( { - 1} \right)}^{1/2 - {m_J}}}\left( {\begin{array}{*{20}{c}}
  			{1/2}&1&{1/2} \\ 
  			{ - {m_J}}&q&{m_J^\prime } 
  	\end{array}} \right)}  \nonumber\\ 
 & \qquad  \times \left\langle {n{P_{1/2}}\left\| {\hat d_q} \right\|{n^\prime }{S_{1/2}}} \right\rangle .
\end{align}

\begin{figure}[t]
	\centering
	\includegraphics[width=0.48\textwidth]{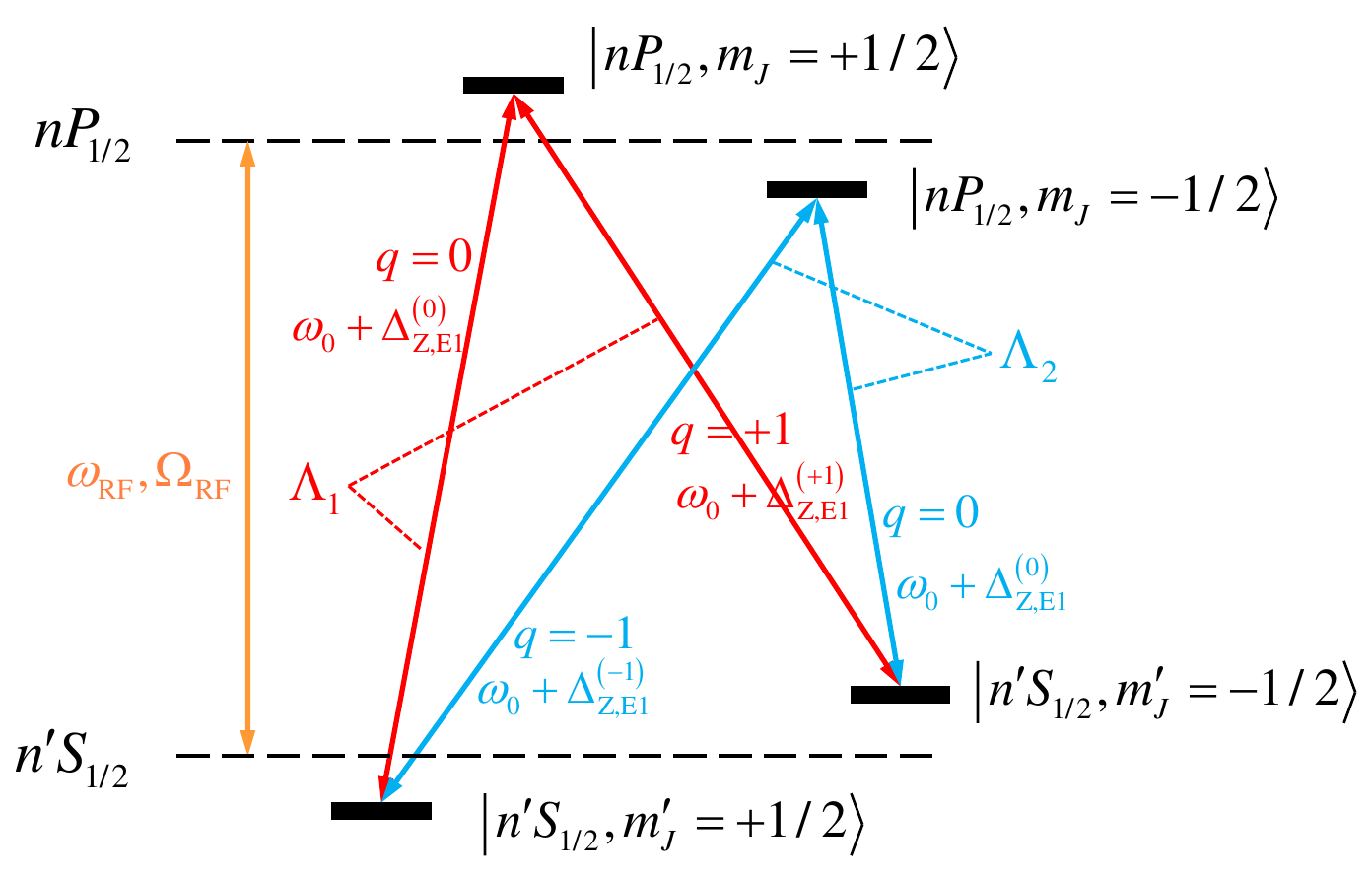}
	\caption{Illustration of pure $\Lambda$ systems.} \label{Lambda_systems}
\end{figure}

\subsection{Pure $\Lambda$ Systems}
In the presence of a weak static magnetic bias field $\mathbf{B}_\text{bias}$ characterized by magnitude $B_\text{bias}$ and orientation $\theta_\text{bias}$, the excited Rydberg states $n^\prime S_{1/2}$ and $nP_{1/2}$ undergo Zeeman splitting into sub-levels $\left\{ {\left| {{n^\prime }{S_{1/2}},m_J^\prime } \right\rangle ,\left| {n{P_{1/2}},{m_J}} \right\rangle } \right\}$. When a pair of optical fields couple these sub-levels to a common excited state in a $\Lambda$ configuration, the atoms are optically pumped into a non-interacting dark superposition, giving rise to Zeeman-resolved EIT resonances whose linewidth and amplitudes depend on the magnetic environment~\cite{QS-52,QS-53}. In this basis, the three spherical components of the RF electric field are mapped onto the relative heights of the Zeeman-resolved EIT peaks.
As sketched in Fig.~\ref{Lambda_systems}, the sub-levels $\left| {{n^\prime }{S_{1/2}},m_J^\prime  =  + 1/2} \right\rangle $ and $\left| {{n^\prime }{S_{1/2}},m_J^\prime  =  - 1/2} \right\rangle $ serve as the $\Lambda$ ground state, while the common excited state is chosen as $ \left| {n{P_{1/2}},{m_J} =  \pm 1/2} \right\rangle  $. The magnetic selection rules of $\Delta m = 0, \pm 1$ then allow a set of three RF-driven transition paths labeled by the polarization index $q = 0, \pm1$. A strong $5S_{1/2} \to 5 P_{3/2}$ pump 
together with an auxiliary repumper optically pumps population into sub-levels $\left\{ {\left| {n^\prime{S_{1/2}} , m^\prime_J} \right\rangle ,\left| {{n }{P_{1/2}} , m_J} \right\rangle } \right\}$. All other hyperfine levels remain far from being resonant and thus are neglected in our system. 


\subsection{E1-Transition Rabi Frequency}
The E1-transition Rabi frequency $\Omega_{{\text{E1}}}$ can be formulated as
\begin{align}\label{Omega_RF}
&  \Omega _{{\text{E1}}}  \left( {\theta_{\text{RF}} } \right)  \nonumber\\
 &= \frac{1}{\hbar }\left| {\left\langle {n{P_{1/2}},{m_J}\left| { - \hat{\mathbf{d}} \cdot \mathbf{E} } \right|{n^\prime }{S_{1/2}},{m^\prime_J}} \right\rangle } \right| \nonumber\\
 &= \frac{{{E_0}}}{\hbar }\left| {\sum\limits_{q =  - 1}^{ + 1} {{\alpha _q}\left( {{\theta _{{\text{RF}}}}} \right){{\left( { - 1} \right)}^{1/2 - {m_J}}}\left( {\begin{array}{*{20}{c}}
				{1/2}&1&{1/2} \\ 
				{ - {m_J}}&q&{{m^\prime_J}} 
		\end{array}} \right)} }\right. 
	\nonumber\\
	& \quad \times \left.{  \left\langle {n{P_{1/2}}\left\| {\hat d_q} \right\|{n^\prime }{S_{1/2}}} \right\rangle } \right|.
\end{align}
Furthermore, the RF field does not directly couple the ground state to the probe state; instead, it collects all second‐order shifts and broadenings of the intermediate state. Physically, all the transition paths contribute a ``self-energy" term ${\Sigma _{{\text{E1}}}}$, which is characterized by
\begin{equation}
	{\Sigma _{{\text{E1}}}} = \sum\limits_{q =  - 1}^{ + 1} {\frac{{{{\left| {\Omega _{{\text{E1}}}^{\left( q \right)}\left( {{\theta _{{\text{RF}}}}} \right)} \right|}^2}}}{{\Delta _{{\text{E1}}}^{\left( q \right)} + \jmath {\gamma _{{\text{RF}}}}}}} ,
\end{equation}
The Rabi frequency of each transition path $q$ is given by
\begin{align}\label{Rabi_freq_E1_q}
	\Omega _{{\text{E1}}}^{\left( q \right)}\left( {{\theta _{{\text{RF}}}}} \right) =& \frac{{{E_0}}}{\hbar }\left| {{\alpha _q}\left( {{\theta _{{\text{RF}}}}} \right){{\left( { - 1} \right)}^{1/2 - {m_J}}}\left( {\begin{array}{*{20}{c}}
				{1/2}&1&{1/2} \\ 
				{ - {m_J}}&q&{{m^\prime_J}} 
		\end{array}} \right) }\right. \nonumber\\ 
	& \times\left. { \left\langle {n{P_{1/2}}\left\| {\hat d_q} \right\|{n^\prime }{S_{1/2}}} \right\rangle } \right|,
\end{align}
and $\gamma_{{\text{RF}}}$ is the dephasing rate associated with the E1-transition driven by the RF field. The corresponding detuning for the transition path $q$ is formulated as 
\begin{equation}\label{Delta_E1_q}
\Delta _{{\text{E1}}}^{\left( q \right)} = {\omega _{{\text{RF}}}} - \left( {{\omega_\text{0,E1}} + \Delta _{{\text{Z,E1}}}^{\left( q \right)}} \right),
\end{equation}
where $\omega_\text{0,E1}$ denotes the natural transition frequency between the $n^\prime S_{1/2}$ and $nP_{1/2}$ states in the absence of external fields. The Zeeman shifit can be expressed as
\begin{equation}\label{Delta_Z_E1_q}
\Delta _{{\text{Z,E1}}}^{\left( q \right)} = \frac{{{\mu _B}{B_{{\text{bias}}}}}}{\hbar }\left( {{g_P}{m_J} - {g_S}m_J^\prime } \right)\cos {\theta _{{\text{bias}}}},
\end{equation}
where $\mu_B$ is the Bohr magneton, and $g_P$, $g_S$ represent the Land\'{e} factors of the $ {n{P_{1/2}}}$ and ${n'{S_{1/2}}}  $ states, respectively.

Note that the RF carrier frequency is typically on the order of  Gigahertz separation away from any magnetic-dipole resonance of the Rydberg manifold, and hence the oscillating RF magnetic field acts predominantly as a time-dependent drive, rather than a quasi-static fluctuation. Its only residual influence on the bare energies is a second-order, off-resonant shift, such as the magnetic AC-Zeeman shift~\cite{QS-64}.  Given the magnitude of the oscillating RF magnetic field $ B_{\text{RF}} \lesssim 10^{-7}~\text{T}$, the resultant level shift is at most a few Hertz, whose orders of magnitude are much smaller than the kiloHertz-scale static Zeeman shift imposed by the bias field. Therefore, we neglect this AC-Zeeman contribution and treat the RF magnetic component solely as a coupling term that mixes the neighboring Zeeman sublevels.

\subsection{Zeeman-Resolved EIT Spectrum}
By accounting for the bias-field-induced Zeeman shift in each transition path, the density matrix element $\rho_{21}^{\text{E1}}$ can be explicitly formulated as~\cite{chen-arxiv}
\begin{equation}\label{rho_21_E1}
\rho _{21}^{{\text{E1}}} = \frac{{\jmath \left( {{\Omega _p}/2} \right)}}{{{\Delta _p} + \jmath {\gamma _{21}} - \frac{{{{\left( {{\Omega _c}/2} \right)}^2}}}{{{\Delta _c} + \jmath {\gamma _{{\text{31}}}} -  \Sigma_{{\text{E1}}} }}}}.
\end{equation}
The above equation encapsulates the components in terms of the RF-dressed self-energy ${\Delta _{{\text{E1}}}^{\left( q \right)}}$ and the Rabi frequency ${\Omega _{{\text{E1}}}^{\left( q \right)}\left( {{\theta _{{\text{RF}}}}} \right)}$ for each transition path $q$.

\begin{figure}[t]
	\centering
	\includegraphics[width=0.45\textwidth]{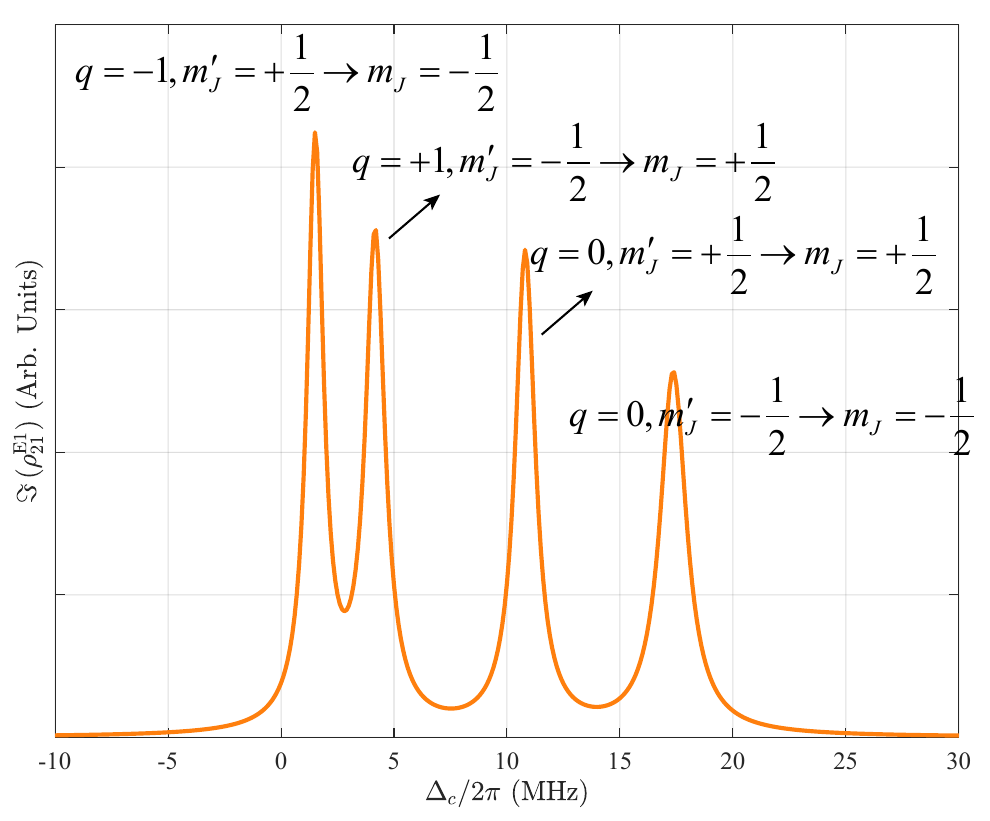}
	\caption{Zeeman-resolved EIT spectrum at fixed $\theta_{\text{RF}} = 30^\circ$ and $\theta_{\text{bias}} = 0^\circ$.} \label{magnetic_EIT}
\end{figure}

\begin{figure*}[t]
	\centering
	\includegraphics[width=0.95\textwidth]{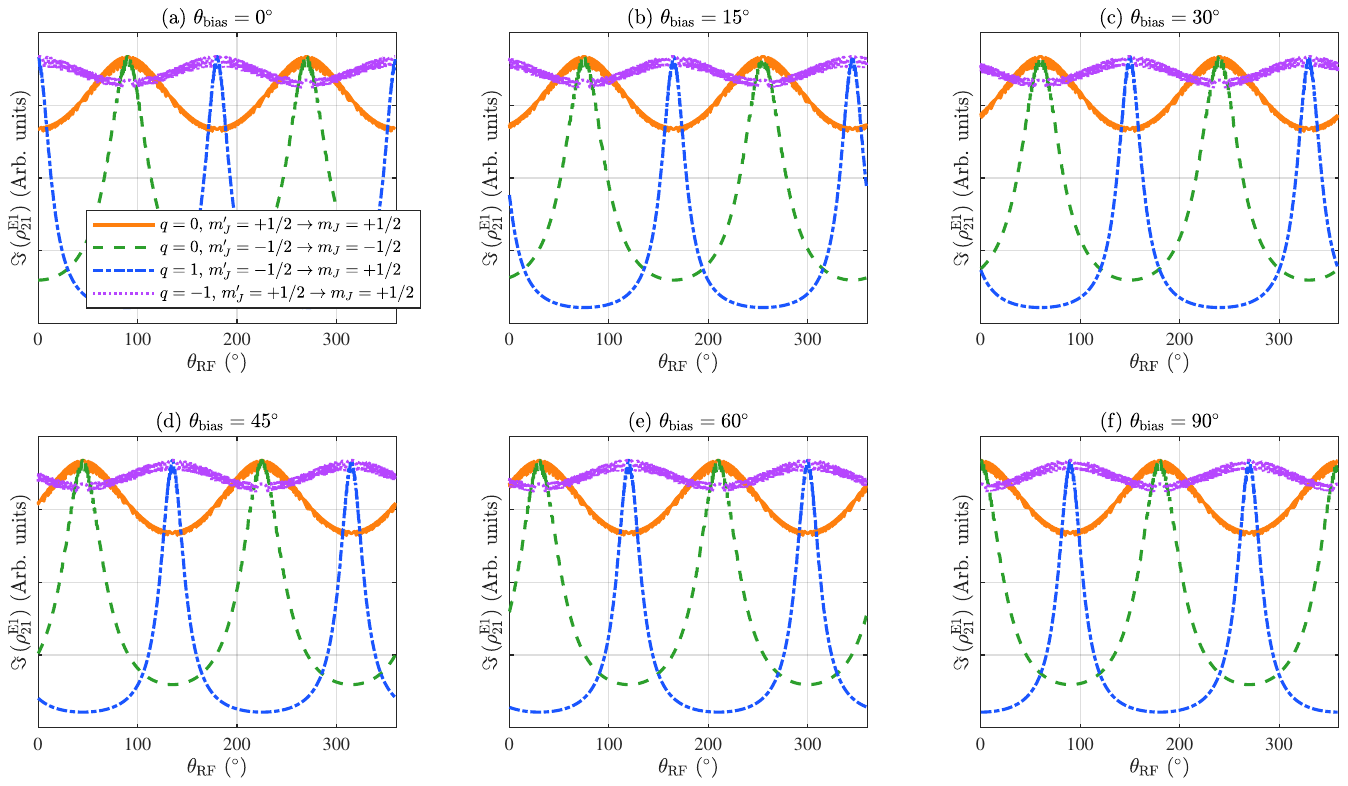}
	\caption{Zeeman-resolved EIT spectrum with varying $\theta_{\text{RF}}$ and $\theta_\text{bias}$.} \label{EIT_vs_2angles_6plots}
\end{figure*}

To validate the analytic expression given in (\ref{rho_21_E1}), simulation results are presented in Figs.~\ref{magnetic_EIT}-\ref{EIT_vs_2angles_6plots} by plotting $\rho _{21}^{\text{E1}}$ curves under different configurations. The atomic and optical parameters are chosen to match recent room-temperature vapor-cell experiments~\cite{chen-arxiv,QS-16}: probe and coupling Rabi frequencies of $\Omega_p /2\pi = 0.04~\text{MHz}$, $\Omega_c /2\pi = 0.67~\text{MHz}$; inverse lifetimes of $\gamma_2 /2\pi = 5.2~\text{MHz}$, $\gamma_3 /2\pi = 3.9~\text{MHz}$, and $\gamma_4 /2\pi = 0.17~\text{MHz}$; as well as zero probe and RF detunings of $\Delta_p = \Delta_{\text{RF}} = 0$. The dipole moment associated with the Rydberg transition is $\hat{d} = -1443.46ea_0$, with $e = 1.6\times10^{19}~\text{C}$ representing the elementary charge and $a_0 = 5.2\times 10^{-11}~\text{m}$ for the Bohr radius. The operation frequency is $6.9~\text{GHz}$. For a near-resonant E1 transition, the natural transition frequency is treated as almost the frequency of the applied RF field, i.e., $\omega_\text{0,E1}/2\pi = 6.9~\text{GHz}$. In the presence of detunings, the actual frequency is given by $\omega_{{\text{RF}}}  = \omega_\text{0,E1} \pm \tilde{\Delta}  $, with $\tilde{\Delta}/ 2\pi \sim 0.8~\text{MHz} $. Given a dephasing rate of $\gamma_{\text{RF}} = 10~\text{kHz}$, a static magnetic bias field of $B_{\text{bias}} = 2~\text{mT}$, an electric-field amplitude of $E_0 = 1~\text{V/m}$, and Land\'{e} factors of $g_S = 2.0$ and $g_P = 0.67$, the Zeeman shifts are on the order of MHz, sufficiently high to lift the degeneracy of the four legitimated transitions.  
Additionally, to extract the polarization (or magnetic field) angle, we rely solely on the relative areas of the Zeeman-resolved AT peaks. As a result, absolute gain calibration is not required. A single reference spectrum acquired at a known angle sets the overall scale, while each subsequent trace is normalized to its simultaneously recorded off-resonant probe transmission.

Specifically, Fig.~\ref{magnetic_EIT} plots $\Im \left( \rho_{21}^{\text{E1}} \right) $ versus the coupling detuning $\Delta_c$. As it transpires, four splitting peaks are observed, whose order and amplitudes coincide with the expected Clebsch-Gordan coefficients and spherical-basis projections. Furthermore, in Fig.~\ref{EIT_vs_2angles_6plots}, we retain the same electric-field amplitude of $E_0 = 1~\text{V/m}$ and sweep the polarization angle $\theta_{\text{RF}}$ over $360^\circ$. Panels (a)-(f) reveal the peak of each Zeeman-resolved transition path for magnetic-field orientations $\theta_{\text{bias}} = 0^\circ, 15^\circ,30^\circ,45^\circ,60^\circ$, and $90^\circ$. As it can be observed, under different $\theta_{\text{bias}}$, their envelopes follow the expected $ \left| \cos \theta_{\text{RF}} \right| $ dependence. Regarding the $\pi$-polarized case, the two mirror transitions, i.e., $q = 0$, differ in absolute magnitude owing to their different Clebsch-Gordan coefficients for $m_J = \pm 1/2$. 
For the $\sigma^\pm$-polarized cases, the maxima occur at $\theta_{\text{RF}} = 90^\circ$ and $\theta_{\text{RF}} = 270^\circ$  when $\theta_{\text{bias}} = 0^\circ$.
Tilting the quantization axis away from the optical-beam direction rotates and re-weights the $\sigma^\pm$ versus $\pi$ admixture: the extrema gradually shift by an amount that scales with $\sin \theta_{\text{bias}}$. 

\section{Detection of $\theta_{B}$}\label{Sec_IV}

Having demonstrated that the RF-field polarization can be retrieved from the characteristics of the EIT spectrum, we now turn our attention to the reconstruction of the RF magnetic-field $\mathbf{B}_{\text{RF}}$. This section carries out the mirror‐image analysis for $\theta_B$'s detection, but on a magnetic-dipole transition basis.

\subsection{Magnetic-Dipole Transition}


We replace the atomic system in (\ref{E1_transition}) with the following four-level system
\begin{equation}\label{M1_transition}
5{S_{1/2}}\mathop  {\longrightarrow} \limits^{\Omega _p } 5{P_{3/2}}\mathop  {\longrightarrow} \limits^{\Omega _c^\prime } n'{P_{1/2}}\mathop  {\longrightarrow} \limits^{ \Omega _{{\text{M1}}} \ } n{P_{3/2}}.
\end{equation}
The fine-structured link $n'{P_{1/2}} \to n{P_{3/2}}$ is electric-dipole disabled but magnetic-dipole enabled, and thus the RF field couples to it only through its magnetic components $\mathbf{B}_{\text{RF}}$.
When imposing a bias field $\mathbf{B}_\text{bias}$, the excited Rydberg states are split into Zeeman sub-levels $\left\{ {\left| {{n^\prime }{P_{1/2}},m_J^\prime } \right\rangle ,\left| {n{P_{3/2}},{m_J}} \right\rangle } \right\}$. Pure $\Lambda$ systems can be constructed, where the two $m^\prime_J = \pm 1/2$ levels of  $P_{1/2} $ serve as the $\Lambda$ ground state and the shared excited state is selected as $\left| {n{P_{3/2}},{m_J} =  \pm 1/2, \pm 3/2} \right\rangle $. 
When $\mathbf{B}_{\text{bias}}$ is rotated away from the quantization $z$-axis, the definitions of the $\pi$ ($\Delta m = 0$) and $\sigma$ ($\Delta m = \pm 1$) couplings are rotated accordingly. This redistributes the population among the Zeeman sub-levels in a manner that encodes the instantaneous orientation of $\mathbf{B}_{\text{bias}}$. In the following, we construct the full Zeeman-resolved Hamiltonian and derive analytical expressions of the M1-transition's Rabi frequency.



\subsection{Magnetic Hamiltonian and M1-Transition Rabi Frequency}
The magnetic Hamiltonian $H_\mu$ can be formulated as
\begin{align}
{H_\mu } =&  - {\hat {\bm{\mu}}} \cdot {{\mathbf{B}}_{{\text{RF}}}} \nonumber\\
=&  - \left\langle {n{P_{3/2}},{m_J}\left| { {\hat {\bm{\mu}}}  \cdot {{\mathbf{B}}_{{\text{RF}}}}} \right|{n^\prime }{P_{1/2}},m_J^\prime } \right\rangle \nonumber\\
=& - {B_{{\text{RF}}}}\sum\limits_{q =  - 1}^{ + 1} {{\beta _q}\left( {{\theta _B}} \right)\left\langle {n{P_{3/2}},{m_J}\left| {{{\hat \mu }_q}} \right|{n^\prime }{P_{1/2}},m_J^\prime } \right\rangle } \nonumber\\
=&  - {B_{{\text{RF}}}}\sum\limits_{q =  - 1}^{ + 1} {\beta _q}\left( {{\theta _B}} \right){{\left( { - 1} \right)}^{3/2 - {m_J}}}\left( {\begin{array}{*{20}{c}}
 			{3/2}&1&{1/2} \\ 
 			{ - {m_J}}&q&{{m^\prime }_J} 
 	\end{array}} \right) \nonumber\\
 & \times \left\langle {n{P_{3/2}}\left\| {{{\hat \mu }_q}} \right\|{n^\prime }{P_{1/2}}} \right\rangle  ,
\end{align}
where the oscillating RF magnetic field $\mathbf{B}_{\text{RF}}$ can be expanded in the same spherical basis $\left( {\bm{\epsilon}}_{-1}, {\bm{\epsilon}}_{0}, {\bm{\epsilon}}_{+1}\right) $, yielding
\begin{equation}
{{\mathbf{B}}_{{\text{RF}}}} = {B_{{\text{RF}}}}\sum\limits_{q =  - 1}^{q =  + 1} {{\beta _q}\left( {{\theta _B}} \right){{\bm{\epsilon}}_q}} ,
\end{equation}
along with 
\begin{equation}
{\beta _0}{\mkern 1mu} \left( {{\theta _B}} \right) = \cos {\theta _B},~{\beta _{ \pm 1}}{\mkern 1mu} \left( {{\theta _B}} \right) =  \mp \frac{{\sin {\theta _B}}}{{\sqrt 2 }}.
\end{equation}
To this end, the M1-transition Rabi frequency can be expressed as
\begin{align}\label{Rabi_freq_M1_q}
&\Omega _{{\text{M1}}} \left( {{\theta _B}} \right) \nonumber\\ 
&=  \frac{{{B_{{\text{RF}}}}}}{\hbar }\left| { \sum\limits_{q =  - 1}^{ + 1} {\beta _q}\left( {{\theta _B}} \right){{\left( { - 1} \right)}^{3/2 - {m_J}}}\left( {\begin{array}{*{20}{c}}
			{3/2}&1&{1/2} \\ 
			{ - {m_J}}&q&{{m^\prime }_J} 
	\end{array}} \right) }\right. \nonumber\\  
& \quad \times \left. {\left\langle {n{P_{3/2}}\left\| {{{\hat \mu }_q}} \right\|{n^\prime }{P_{1/2}}} \right\rangle } \right|.
\end{align}
Likewise, the self-energy term for the M1-transition, namely ${\Sigma _{{\text{M1}}}}$, which collects the contributions from all allowed transition paths, can be written as
\begin{equation}
{\Sigma _{{\text{M1}}}} = \sum\limits_{q =  - 1}^{ + 1} {\frac{{{{\left| {\Omega _{{\text{M1}}}^{\left( q \right)}\left( {{\theta _B}} \right)} \right|}^2}}}{{\Delta _{{\text{M1}}}^{\left( q \right)} + \jmath \gamma _{{\text{RF}}}^\prime }}} .
\end{equation}
The Rabi frequency $\Omega _{{\text{M1}}}^{\left( q \right)}\left( {{\theta _B}} \right)$ for the transition path $q$ is in the form of
\begin{align}
\Omega _{{\text{M1}}}^{\left( q \right)}\left( {{\theta _B}} \right)  
= & \frac{{{B_{{\text{RF}}}}}}{\hbar }\left| { {\beta _q}\left( {{\theta _B}} \right){{\left( { - 1} \right)}^{3/2 - {m_J}}}\left( {\begin{array}{*{20}{c}}
			{3/2}&1&{1/2} \\ 
			{ - {m_J}}&q&{{m^\prime }_J} 
	\end{array}} \right) }\right. \nonumber\\  
& \times \left. {\left\langle {n{P_{3/2}}\left\| {{{\hat \mu }_q}} \right\|{n^\prime }{P_{1/2}}} \right\rangle } \right|,
\end{align}
in which ${\gamma _{{\text{RF}}}^\prime }$ denotes the dephasing rate associated with the M1-transition driven by the RF field. The corresponding detuning for the transition path $q$ is given by
\begin{equation}
\Delta _{{\text{M1}}}^{\left( q \right)} = {\omega _{{\text{RF}}}} - \left( {{\omega_\text{0,M1}} + \Delta _{{\text{Z,M1}}}^{\left( q \right)}} \right).
\end{equation}
Consequently, the Zeeman shift becomes
\begin{equation}
\Delta _{{\text{Z,M1}}}^{\left( q \right)} = \frac{{{\mu _B}{B_{{\text{bias}}}}}}{\hbar }\left( {{g_P}{m_J} - g^\prime _P m_J^\prime } \right)\cos {\theta _{{\text{bias}}}},
\end{equation}
where ${{g_P}}$ and ${g^\prime_P}$ gives the Land\'{e} factors for  $nP_{3/2}$ and $n^\prime P_{1/2}$, respectively. Therefore, we obtain the Zeeman-resolved EIT spectrum by analyzing the density matrix element of
\begin{equation}\label{rho_21_M1}
\rho _{21}^{{\text{M1}}} = \frac{{\jmath \left( {\Omega _p /2} \right)}}{{\Delta _p  + \jmath \gamma _{21}^\prime  - \frac{{{{\left( {\Omega _c^\prime /2} \right)}^2}}}{{\Delta _c^\prime  + \jmath \gamma _{31}^\prime  - \Sigma_{{\text{M1}}} }}}},
\end{equation}
where we use a prime superscript ($^\prime$) to distinguish these variables from those appearing in (\ref{rho_21_E1}) of the E1-transition.

\subsection{Reconstruction of $\theta_{B}$ From Bias-Field Scanning}
Having derived the M1-transition Rabi frequency and the corresponding Zeeman-resolved EIT spectrum, we now discuss how a controlled rotation of the static bias field enables a full reconstruction of $\theta_B$. The procedure consists of two deterministic steps.
\begin{enumerate}
\item Let ${{\hat n}_i}\left( {i = 1,2,...} \right)$ be a set of bias-field orientations produced by a three-axis Helmholtz coil. For each orientation ${{\hat z}_i}$, the spherical basis $ {\left\{ {{\bm{\epsilon}}_q^{\left( i\right)} } \right\}_{q = 0, \pm 1}} $ is obtained by a passive rotation of the standard $ \left( x,y,z \right)  $ basis into $\hat{n}_i$.
The M1-transition coupling associated with the $\Delta m=q$ Zeeman channel is determined by the spherical-basis component of the RF magnetic field with respect to the quantization axis $\hat{n}_i$. Specifically, we define
\begin{equation}
	B^{\left( i\right) }_{\text{RF},q} = \left|\mathbf{B}_{\text{RF}}\cdot\boldsymbol{ \epsilon }^{\left( i\right)}_{q}\right|, q\in\{0,\pm1\},
\end{equation}
which corresponds to the $\pi$ ($q=0$) and $\sigma^\pm$ ($q=\pm1$) components, respectively.


\item Selecting two mutually orthogonal bias axes, e.g., $\hat{n_1} \parallel z$ and $\hat{n_1} \parallel y$, furnishes an invertible system
\begin{align}\label{B_RF_equation}
\left[ \begin{gathered}
	{\bm{\epsilon}}_{ + 1}^{\left( 1 \right),T} \hfill \\
	{\bm{\epsilon}}_0^{\left( 1 \right),T} \hfill \\
	{\bm{\epsilon}}_{ - 1}^{\left( 1 \right),T} \hfill \\
	{\bm{\epsilon}}_{ + 1}^{\left( 2 \right),T} \hfill \\
	{\bm{\epsilon}}_0^{\left( 2 \right),T} \hfill \\
	{\bm{\epsilon}}_{ - 1}^{\left( 2 \right),T} \hfill \\ 
\end{gathered}  \right]{{\mathbf{B}}_{{\text{RF}}}} = \left[ \begin{gathered}
	\mathfrak{b}_{ + 1}^{\left( 1 \right)}B_{{\text{RF,}} + 1}^{\left( 1 \right)} \hfill \\
	\mathfrak{b}_0^{\left( 1 \right)}B_{{\text{RF,0}}}^{\left( 1 \right)} \hfill \\
	\mathfrak{b}_{ - 1}^{\left( 1 \right)}B_{{\text{RF,}} - 1}^{\left( 1 \right)} \hfill \\
	\mathfrak{b}_{ + 1}^{\left( 2 \right)}B_{{\text{RF,}+1}}^{\left( 2 \right)} \hfill \\
	\mathfrak{b}_0^{\left( 2 \right)}B_{{\text{RF,0}}}^{\left( 2 \right)} \hfill \\
	\mathfrak{b}_{ - 1}^{\left( 2 \right)}B_{{\text{RF,}} - 1}^{\left( 2 \right)} \hfill \\ 
\end{gathered}  \right],
\end{align}
where ${\mathfrak{b}}_q^{\left( i\right) }$ encodes the $0 / \pi$ phase (sign) of the complex spherical component $\tilde B_{\text{RF},q}^{\left( i \right)}  = \mathbf{B}_\text{RF} \cdot {\boldsymbol{\epsilon}}_q^{\left( i \right)} $. In practice, ${\mathfrak{b}}_q^{\left( i\right) }$ can be obtained by a phase-sensitive (I/Q) readout referenced to the RF source.
Then, the oscillating RF magnetic field $\mathbf{B}_\text{RF}$ can be obtained by solving equation~(\ref{B_RF_equation}).

\end{enumerate}

In our single vapor cell platform considered, the electric‐ and magnetic‐field measurements are conducted back‐to‐back by rapid frequency retuning, with no mechanical changes to the vapor cell, bias field, or beam alignment. More specifically, first, the probe laser is locked to  $5 S_{1/2} \to 5 P_{3/2}$ and the coupling laser to $5 P_{3/2} \to n^\prime S_{1/2}$, while the RF source is tuned to the on-resonance E1-transition $n^\prime S_{1/2} \to nP_{1/2}$. The resultant Zeeman-resolved EIT spectrum directly yields the polarization angle $\theta_{\text{RF}}$. Subsequently, within the same interleaved measurement sequence, an acousto-optic modulator (AOM) shifts the coupling laser in order to address the transition $5 P_{3/2} \to n^\prime P_{1/2}$, and the RF carrier is retuned from $\omega_{\text{0,E1}}$ to $\omega_{\text{0,M1}}$ for driving an M1-allowed transition $n^\prime P_{1/2} \to nP_{3/2}$. The prompt electronic switching ensures that both measurements share an effectively identical geometry on a timescale much shorter than experimental drifts, enabling a joint reconstruction of $\left( {\theta_\text{RF}, \theta_B}\right) $ without moving parts or beam realignment.


\section{Quantum Cram\'{e}r-Rao Bound (QCRB)}\label{Sec_V}

The preceding sections established how the joint electric- and magnetic-field information is characterized on the Zeeman-resolved EIT spectrum. We now ask a more fundamental question: What is the ultimate precision with which those field parameters can be inferred, regardless of the particular measurement strategy? In quantum estimation theory, this limit is set by the quantum Cram\'{e}r-Rao bound, which in turn is determined by the quantum Fisher information (QFI) or, in the multi-parameter case, the QFIM~\cite{QS-61,QS-62}. This metric is introduced to quantify how far quantum resources can push interferometric sensitivity beyond the SQL~\cite{2004_science}. It serves as an absolute benchmark against which any practical readout scheme, such as optical, microwave or hybrid, must be evaluated. Therefore,  we derive the QCRB in this section to examine its theoretical limit.

\subsection{Theoretical QCRB}

The angle pair $\left( {{\theta _{{\text{RF}}}},{\theta _B}} \right)$ is extracted from two sequential, statistically independent spectra that are measured within the same experimental cycle. Each parameter of interest, $\theta_{\text{RF}}$ and $\theta_B$, is embedded in its respective density matrix $\bm{\rho }\left( {{\theta _{{\text{RF}}}}} \right)$ and $\bm{\rho }\left( {{\theta _B}} \right)$. These two states collaboratively form  a direct-sum state that characterizes the overall measurement
\begin{equation}
\bm{\rho }\left( {{\theta _{{\text{RF}}}},{\theta _B}} \right) = \bm{\rho }\left( {{\theta _{{\text{RF}}}}} \right) \oplus \bm{\rho }\left( {{\theta _B}} \right).
\end{equation}
Since the two blocks occupy orthogonal subspaces, the symmetric logarithmic derivatives act independently in each block, and the global QFIM is simply the direct sum of the single-parameter Fisher information, which is formulated as
\begin{equation}
{\mathbf{F}} = \left[ {\begin{array}{*{20}{c}}
		{{{\mathbf{F}}_{{\theta _{{\text{RF}}}}}}}&0 \\ 
		0&{{{\mathbf{F}}_{{\theta _B}}}} 
\end{array}} \right] \in {\mathbb{R}^{2 \times 2}}.
\end{equation}
As established by Theorem~1 in~\cite[Eq.~(4)]{QS-61}, the element ${{\mathbf{F}}_{{\theta _i}}}$ is given in~(\ref{QFIM}), shown at the top of the next page. 
\begin{figure*}[!t]
\begin{equation}\label{QFIM}
{{\mathbf{F}}_{{\theta _i}}} = 2{\text{vec}}{\left[ {\frac{\partial }{{\partial {\theta _i}}}\bm{\rho }\left( {{\theta _i}} \right)} \right]^H}{\left[ {{\bm{\rho }^*}\left( {{\theta _i}} \right) \otimes {\mathbf{ I }} + {\mathbf{I}} \otimes \bm{\rho }\left( {{\theta _i}} \right)} \right]^{ - 1}}{\text{vec}}\left[ {\frac{\partial }{{\partial {\theta _i}}}\bm{\rho }\left( {{\theta _i}} \right)} \right],{\theta _i} \in \left\{ {{\theta _{{\text{RF}}}},{\theta _B}} \right\}.
\end{equation}
\hrule
\end{figure*}
We then arrive at the QCRB, which is formulated as
\begin{equation}\label{QCRB}
{\text{QCRB}}\left( {{{\hat \theta }_i}} \right) \ge \frac{1}{\nu }{{\mathbf{F}}^{ - 1}},{\hat \theta _i} \in \left\{ {{{\hat \theta }_{{\text{RF}}}},{{\hat \theta }_B}} \right\},
\end{equation}
where $\nu$ denotes the number of experimental repetitions, representing the number of independent and identically distributed (i.i.d.) measurement outcomes contributing to the statistical average. Since the QCRB yields a lower bound on the variance, its square root can be treated as the achievable angular resolution, given by $\sqrt {\frac{1}{\nu }{{\mathbf{F}}^{ - 1}}} $, with units of radians. This analytic paradigm exhibits a dual but independent dependence on both $\theta_{{\text{RF}}}$ and $\theta_B$. Hence the QFIM determines the joint quantum-enhanced precision for a single-cell DoA reconstruction, without the need for mechanical reconfiguration, spatial diversity, or external phase referencing.

\subsection{SNR-Limited QCRB}

In practical measurements, the DoA is inferred by analyzing the EIT spectrum, where the results are inevitably corrupted by various noise sources. Inspired by the classical CRB behavior in ULA-based receivers, where the CRB scales inversely with the SNR, we now introduce an SNR definition tailored to the QFIM. Given that the parameter dependence of the steady‐state density matrix ${\bm{\rho}}\left( { \theta_i } \right)$ carries the information we expect to extract, we therefore define the sensitivity operator ${{\mathbf{A}}_{{\theta _i}}} = \frac{\partial }{{\partial {\theta _i}}}{\bm{\rho}}\left( {{\theta _{{\text{RF}}}},{\theta _B}} \right)$, whose signal power is naturally given by its squared Frobenius norm $\left\| {{{\mathbf{A}}_{{\theta _i}}}} \right\|_F^2{\text{ = Tr}}\left[ {{\mathbf{A}}_{{\theta _i}}^H{{\mathbf{A}}_{{\theta _i}}}} \right]$. In this case, the QFIM in (\ref{QFIM}) can be trimmed to
\begin{align}
{{\mathbf{F}}_{\theta_i}} = & 2{\left[ {{\text{vec}}\left( {{{\mathbf{A}}_{{\theta _i}}}} \right)} \right]^H}{\left[ {{{\bm{\rho}}^*}\left( { \theta_i } \right) \otimes {\mathbf{I}} + {\mathbf{I}} \otimes {\bm{\rho}}\left( { \theta_i } \right)} \right]^{ - 1}} \nonumber\\
&\times {\text{vec}}\left( {{{\mathbf{A}}_{{\theta _i}}}} \right).
\end{align}
When the density matrix ${\bm{\rho}}\left( { \theta_i } \right)$ exhibits slight variations within the given range of $\theta_i$, or the atomic system is coupled by the weak probe laser, the following approximation holds
\begin{equation}\label{rho_approx}
{{\bm{\rho}}^*}\left( {\theta_i } \right) \otimes {\mathbf{I}} + {\mathbf{I}} \otimes {\bm{\rho}}\left( {\theta_i } \right) \approx 2{\mathbf{I}_{16}}.
\end{equation}
The proof of (\ref{rho_approx}) is given in Appendix~\ref{appendix_proof_rho}. Consequently, the QFIM $\mathbf{F}$ can be approximated as
\begin{equation}\label{QFIM_approx}
{\mathbf{F}} \approx 4\left[ {\begin{array}{*{20}{c}}
		{\left\| {{{\mathbf{A}}_{{\theta _{{\text{RF}}}}}}} \right\|_F^2}& 0 \\ 
		0 &{\left\| {{{\mathbf{A}}_{{\theta _B}}}} \right\|_F^2} 
\end{array}} \right].
\end{equation}
Upon denoting the total noise by $\sigma^2_{\text{total}}$, we can specify the SNR as
\begin{equation}\label{SNR}
{\text{SN}}{{\text{R}}_{{\theta _i}}} = \frac{{\left\| {{{\mathbf{A}}_{{\theta _i}}}} \right\|_F^2}}{{\sigma _{{\text{total}}}^2}},{\theta _i} \in \left\{ {{\theta _{{\text{RF}}}},{\theta _B}} \right\}. 
\end{equation}

\begin{remark}
Note that in (\ref{SNR}), ${{\mathbf{A}}_{{\theta _i}}} = \frac{\partial }{{\partial {\theta _i}}}{\bm{\rho}}\left( {{\theta _{{\text{RF}}}},{\theta _B}} \right)$ inherits the electric-field amplitude $E_0$ via the Rabi frequency, i.e., ${\Omega _{{\text{E1}}}} \propto {E_0}$. Hence, $\left\| {{{\mathbf{A}}_{{\theta _i}}}} \right\|_F^2$ already scales with the received power encountered by the Rydberg atomic receiver. The QFIM-tailored SNR here is a measurement-agnostic form of ``slope-squared over noise". Under the weak-response approximation and the chain rule from $\theta_i$ to $\Omega_\text{E1}$ and $E_0$, it reduces to the implementation-level expressions presented in our earlier work~\cite{chen-arxiv}. For example, regarding the LO-dressed Rydberg atomic receiver, ${{{\mathbf{A}}_{{\theta _i}}}} \propto  \kappa^2 \left( \wp_\text{RF}^2 / \hbar^2 \right) P_\text{Rx} \left| h\right|^2 $, such that ${\text{SN}}{{\text{R}}_{{\theta _i}}}$ coincides with~\cite{chen-arxiv} after identifying the optical-electronic gains and adopting the same noise sources.

\end{remark}

Then, upon substituting (\ref{SNR}) into (\ref{QFIM_approx}), we obtain 
\begin{equation}\label{QFIM_SNR}
{\mathbf{F}} \approx 4\sigma _{{\text{total}}}^2\left[ {\begin{array}{*{20}{c}}
		{{\text{SN}}{{\text{R}}_{{\theta _{{\text{RF}}}}}}}& 0 \\ 
		0 &{{\text{SN}}{{\text{R}}_{{\theta _B}}}} 
\end{array}} \right].
\end{equation}
We thus arrive at the QCRB as a function of SNR. Additionally, the total noise power $\sigma_{{\text{total}}}^2$ includes contributions from both external and internal noise sources~\cite{QS-32,QS-34,chen-arxiv}. The external noise is primarily from blackbody radiation~\cite{QS-32}, while the internal noise comprises a diverse range of sources, including quantum projection noise, photon shot noise, photodetector noise, and thermal noise. For exactly modeling these noise components, the motivated readers might like to refer to the discussions in~\cite{QS-34,chen-arxiv,terry-trans}.

\section{Simulation Results}\label{Sec_VI}
\subsection{Configuration}

In this section, we present simulation results for characterizing the DoA‐estimation precision achieved by a single Rydberg atomic receiver. Most optical and atomic parameters are consistent with those used in Figs.~\ref{magnetic_EIT}-\ref{EIT_vs_2angles_6plots}, but we incorporate several deliberate adjustments. All values are based on well-established spectroscopic data to ensure consistency with~\cite{QS-16,QS-51,QS-53,chen-arxiv}.
Regarding the E1-transition, the probe and coupling Rabi frequencies are set to $\Omega_{p}/2\pi  = 0.04~\text{MHz}$ and $\Omega_{c}/2\pi = 0.67~\text{MHz}$, respectively. The inverse lifetimes of states $|2\rangle$, $|3\rangle$, and $|4\rangle$ are given by $\gamma_{2}/2\pi = 5.2~\text{MHz}$, $\gamma_{3}/2\pi = 3.9~\text{MHz}$, and $\gamma_{4}/2\pi = 0.17~\text{MHz}$, so that $\Omega_{p}/\gamma_{21} \sim 10^{-2}$, and the approximation in Eq.~(\ref{rho_approx}) holds with an accuracy better than 1\%. We assume zero detuning for the probe laser, i.e., $\Delta_p = 0$.  The dipole of the E1-transition is $\hat{d} = -1443.46ea_0$, and for the $\left| 1 \right\rangle  \to \left| 2 \right\rangle $ transition, the dipole moment is  $\left( 2.5ea_0\right) ^2$, with $e = 1.6\times10^{-19}~\text{C}$ representing the elementary charge and $a_0 = 5.29\times 10^{-11}~\text{m}$ for the Bohr radius. 
In the E1 step, the RF carrier is set to $\omega_{\text{0,E1}}/ 2\pi \approx 6.9~\text{GHz}$ (near resonance of the selected E1 Rydberg transition).
In the presence of detunings, the actual carrier frequency becomes $\omega_{{\text{RF}}}  = \omega_{\text{0,E1}} \pm \tilde{\Delta}  $, where $\tilde{\Delta}/ 2\pi \sim 0.8~\text{MHz} $. The static bias field is $B_\text{bias} = 2~\text{mT}$, with Land\'{e} factors $g_{S}=2.0$, $g_{P}=0.67$, and the electric-field amplitude is $E_0 = 1~\text{V/m}$.

As regards to the M1-transition, all optical powers, decay rates and the static bias field are kept identical to the E1-transition measurement so that the two spectra can be acquired sequentially within a single experimental cycle. The carrier frequency is hopped to  $\omega_{\text{0,M1}} $, which is chosen to lie in the same GHz band for the selected M1-allowed Rydberg transition.
The coupling Rabi frequency is retuned to $\Omega_{c}^\prime/2\pi = 0.8~\text{MHz}$, respectively. The Land\'{e} factors for $n^\prime P_{1/2}$ and $nP_{3/2}$ are given by $g_{S}=0.67$ and $g_{P}=1.33$, respectively. 
Our model assumes a $1~\text{cm}$ vapor cell filled with ground-state atoms at a density of $4.89\times10^{16}~\text{m}^{-3}$, held at room temperature $290~\text{K}$. The two back-to-back E1/M1 spectra are acquired using prompt electronic switching (RF frequency hopping and AOM shifting) on a timescale much shorter than experimental drifts, so that the geometry can be regarded as effectively unchanged during one interleaved measurement sequence.
We parameterize the simulations by the local incident RF field at the vapor cell, characterized by $E_0$.
When desired, $E_0$ can be mapped to a transmitter power and distance via standard propagation conditions and antenna gain. However, our DoA reconstruction and QCRB analysis are formulated in an input-referred manner and therefore do not rely on a specific link distance. Typical vapor-cell experiments operate at centimeter-to-decimeter antenna-cell separations~\cite{QS-11,QS-9-TAP-2025}.
From these parameters, one can compute an equivalent effective aperture for the Rydberg receiver, i.e.,
\begin{equation}\label{A_eff}
{A_{{\text{eff}}}} = \frac{{4\pi {Z_0}N_{\text{atoms}}{{\hat d}^2}{\omega _{{\text{RF}}}}{T_2}}}{\hbar },
\end{equation}
whose physical meaning reflects the quantum-mechanical energy exchange efficiency between the atoms and the incident RF field. One can refer to~\cite[Appendix B]{chen-arxiv} for detailed derivations. All density‐matrix evolutions and spectrum calculations were carried out using the QuTiP toolkit~\cite{QuTip}, including the full Zeeman sub-level structure of the four‐level system considered.

\subsection{QCRB Performance}

\begin{figure}[t]
	\centering
	\includegraphics[width=0.43\textwidth]{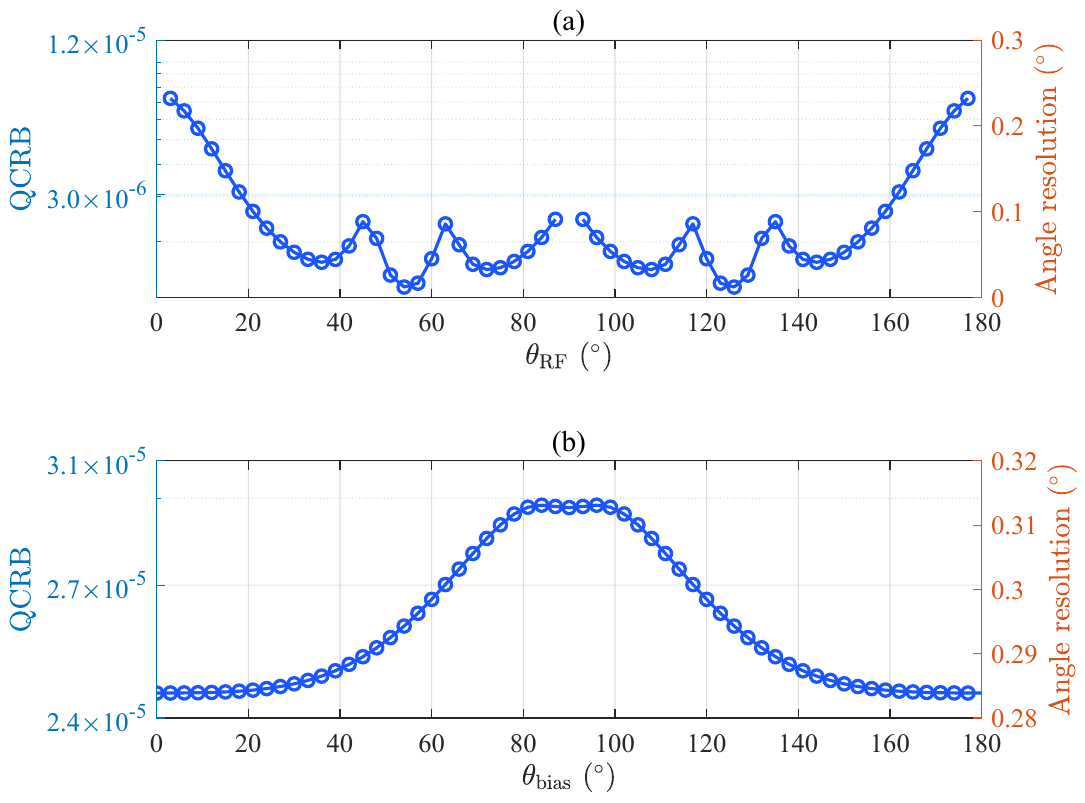}
	\caption{QCRB versus (a) the polarization angle $\theta_{\text{RF}}$ when $\theta_\text{bias} = 45^\circ$ and (b) the orientation of the bias field $\theta_\text{bias}$ when $\theta_{\text{RF}} = 30^\circ$.} \label{QCRB_vs_thetaB_and_thetaRF}
\end{figure}

In this subsection, we examine the QCRB performance versus the polarization angle $\theta_{\text{RF}}$ and the bias-field orientation $\theta_\text{bias}$. Given the approximation in~(\ref{QFIM_approx}), the QFIM is diagonal and yields two distinct marginal bounds, i.e.,  $\text{QCRB} \left( {\theta_{\text{RF}}; \theta_{{\text{bias}}} } \right) \ge 1/ \left( \nu{\mathbf{F}}_{11} \right) $ and $\text{QCRB} \left( { \theta_{B}; \theta_{{\text{bias}}}} \right) \ge 1/ \left( \nu \mathbf{F}_{22}\right) $. A dual-$y$ axis is used for visualization: the left $y$-axis shows the QCRB performance, while the right $y$-axis gives the corresponding angle resolution in degrees. In Fig.~\ref{QCRB_vs_thetaB_and_thetaRF}(a), we plot $\text{QCRB}_{\theta_{\text{RF}}}$ by sweeping $\theta_{\text{RF}}$ with the bias-field orientation fixed at $\theta_{\text{bias}}=45^\circ$.
More specifically, the QCRB curve is strictly mirror-symmetric about $\theta_{\text{RF}}$ and truncated at $\theta_{\text{RF}} = 0^\circ, 90^\circ, 180^\circ$, because the QCRB explodes to $10^{10}$ there. The reason for this pronounced divergence lies in a pair of  extreme polarization conditions. (i) $\theta_{\text{RF}} = 0^\circ, 180^\circ$: the RF field is parallel to the normal of the reference plane and carries only the $\pi$ component; (ii)  $\theta_{\text{RF}} = 90^\circ$: the RF field lies inside the reference plane and carries only the $\sigma^\pm$ components. For any case, only a single Zeeman component ($\pi$, $\sigma^+$ or $\sigma^-$) is driven. As a result, at least one of the Zeeman-resolved transitions (i.e., $q=0$, $q=\pm1$) carries zero dipole strength, causing the Fisher information ${\mathbf{F}} \propto {{\mathbf{A}}_{{\theta _i}}}$ to approach zero, and consequently, the QCRB diverges toward infinity. Once the polarization deviates from these ``forbidden" angles, the four transition paths begin to contribute, and the information content recovers rapidly. This gives rise to a series of local minima in the QCRB approximately every $\sim 20^\circ$, each corresponding to a locally optimal angle resolution of $\sim 0.02^\circ$.

\begin{figure}[t]
	\centering
	\includegraphics[width=0.48\textwidth]{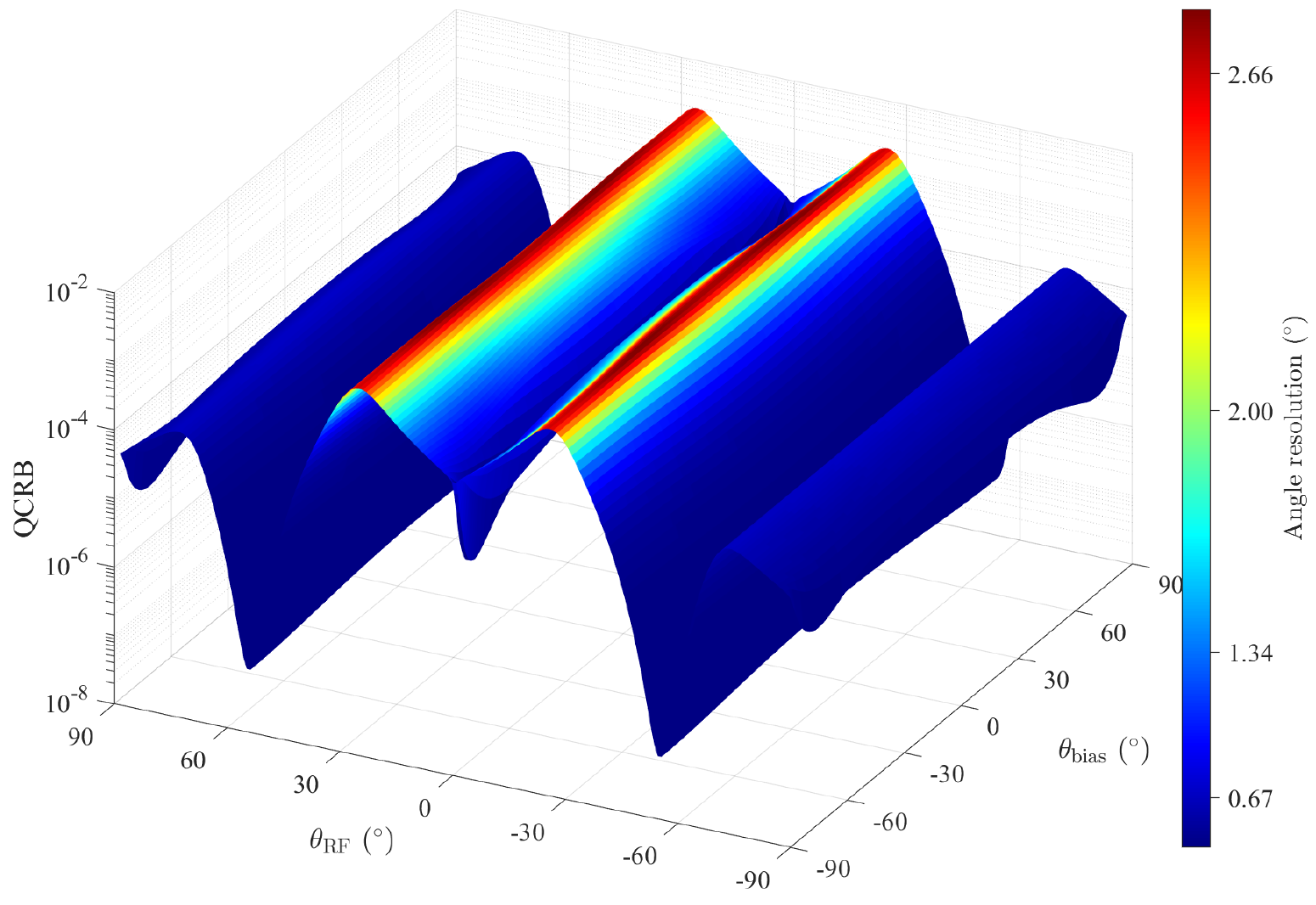}
	\caption{QCRB landscape in the $\theta_{\text{RF}}-\theta_\text{bias}$ plane.} \label{QCRB_vs_thetaB_and_thetaRF_3d}
\end{figure}

With the polarization angle fixed at $\theta_{\text{RF}} = 30^\circ$, the curve in Fig.~\ref{QCRB_vs_thetaB_and_thetaRF}(b) is almost double-peaked. The best angle resolution occurs when $\theta_\text{bias} = 0^\circ, 180^\circ$, i.e., when $\mathbf{B}_\text{bias}$ is orthogonal to the polarization plane and the quantization axis remains close to the laboratory $z$-axis. As a result, the transmission gradients of all transition paths are superimposed within the detection plane, and the Fisher information is hence maximized. 
Rotating the field towards $\theta_\text{bias} \approx 90^\circ$  redistributes the polarization weights, compresses the slope, and hence degrades the Fisher information by roughly an order of magnitude.

\begin{figure}[t]
	\centering
	\includegraphics[width=0.5\textwidth]{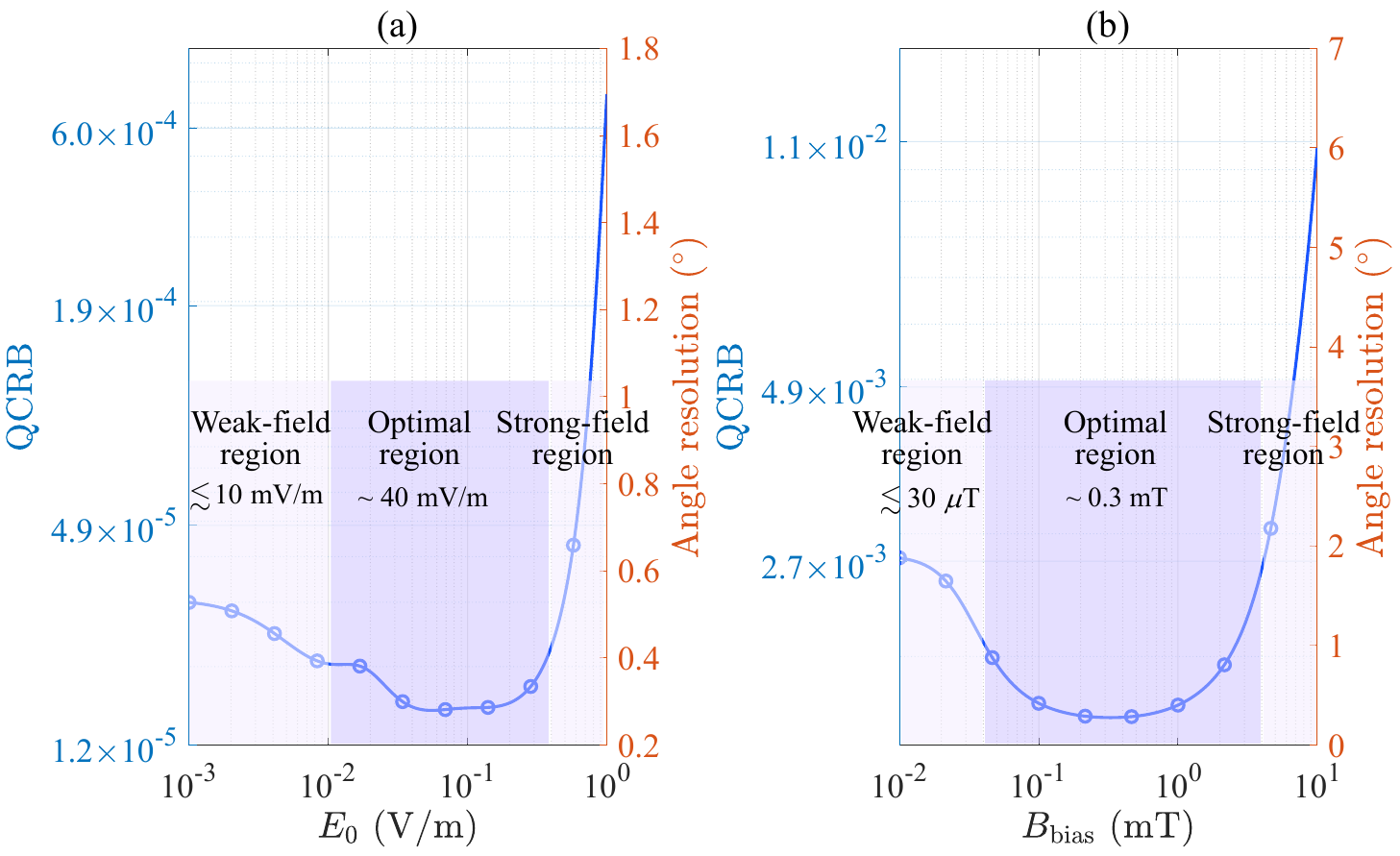}
	\caption{QCRB versus (a) the electric-field amplitude $E_0$ when $B_\text{bias} = 0.2~\text{mT}$ and (b) the static bias field $B_\text{bias}$ when $E_0 = 0.1~\text{V/m}$.} \label{QCRB_vs_E_and_B}
\end{figure}

\begin{figure}[t]
	\centering
	\includegraphics[width=0.47\textwidth]{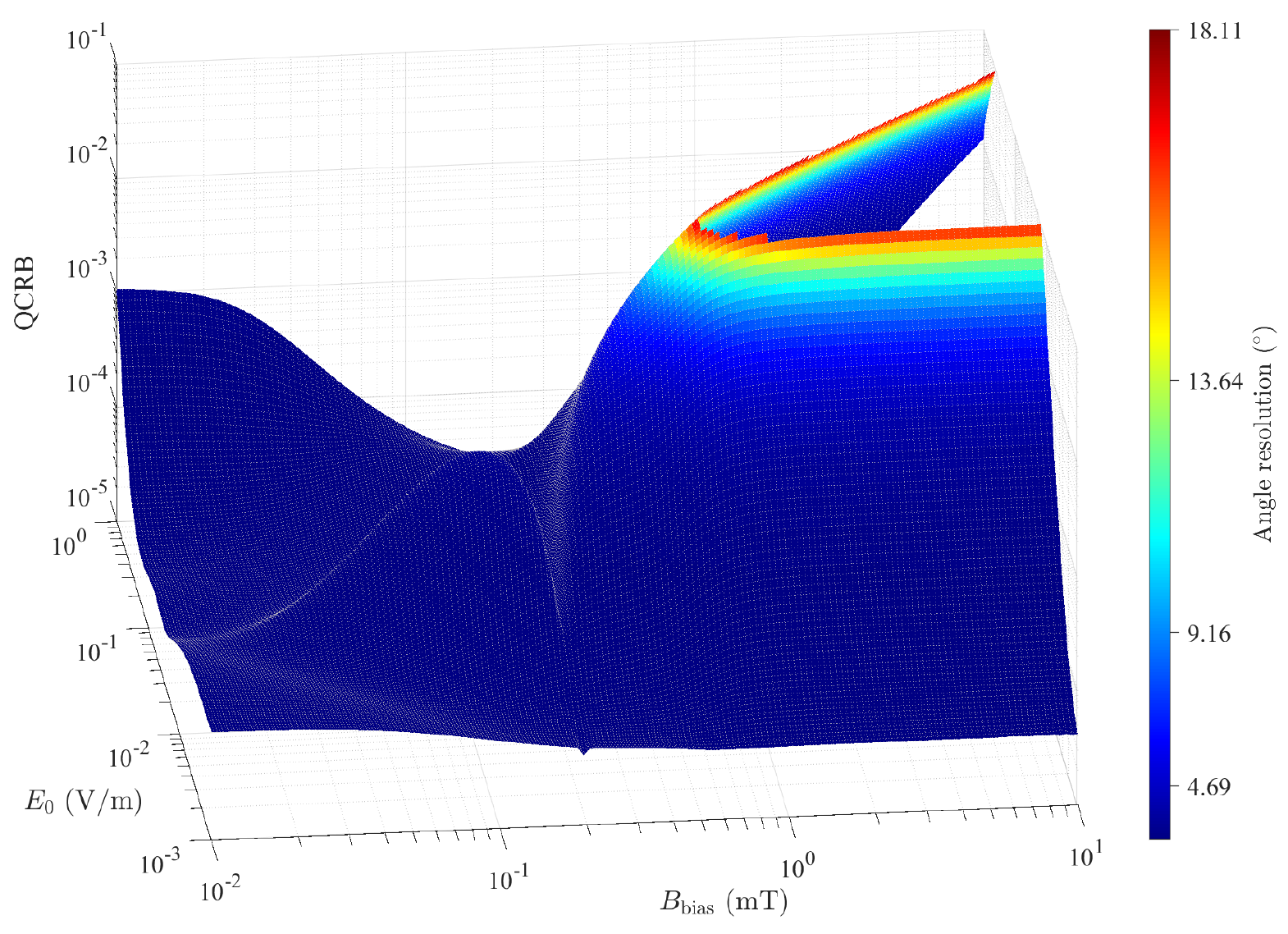}
	\caption{QCRB landscape in the $E_0-B_\text{bias}$ plane.} \label{QCRB_vs_E_and_B_3d}
\end{figure}

Specifically, Fig.~\ref{QCRB_vs_thetaB_and_thetaRF_3d} presents the three-dimensional QCRB surface over the range of $\theta_{\text{RF}}\in \left[ -90^{\circ},+90^{\circ}\right] $ and $\theta_\text{bias}\in\left[ -90^{\circ},+90^{\circ}\right] $, visualizing the joint-estimation QCRB performance for the angle pair $ \left( \theta_{\text{RF}}, \theta_{B} \right) $. A pair of pronounced ridges, arising at $\theta_{\text{RF}}=0^{\circ},\pm90^{\circ}$, mark the Fisher-information zeros already inferred from Fig.~\ref{QCRB_vs_thetaB_and_thetaRF}(a). Furthermore, a narrow valley extends along $\theta_{\text{RF}}\approx \pm45^{\circ}$; its floor reaches the global minimum ($\sim 0.02^{\circ}$) when the magnetic field is orthogonal to the polarization plane (i.e., $\theta_\text{bias}\approx0^{\circ},180^{\circ}$) and gradually rises by an order of magnitude as the field is rotated towards $\theta_\text{bias}\approx90^{\circ}$. This landscape thus provides a global view: staying away from the polarization ``null" ridges ensures operation with the low-lying valley, where sub-$0.05^{\circ}$ quantum-limited angle resolution is achieved.

In Fig.~\ref{QCRB_vs_E_and_B}(a), the static bias field is fixed at $B_\text{bias} = 0.2~\text{mT}$, while the electric-field $E_0$ is swept; in Fig.~\ref{QCRB_vs_E_and_B}(b), $E_0$ is fixed at $0.1~\text{V/m}$, while $B_\text{bias}$ is varied. Both sweeps reveal the three-regime structure. (i) \textit{Weak-field region:} For ${E_0} \lesssim 10~\text{mV/m} $ or $B_\text{bias} \lesssim 30~\mu\text{T}$, the Autler-Townes doublet is obscured by the linewidth, the slope of the probe transmission spectrum with respect to the incident angle approaches zero, and the QFIM collapses. (ii) \textit{Optimal region:} When the dressed-state splitting matches the composite linewidth, the differential phase shift per unit angle is maximized, while the quantum projection noise remains sub-dominant. Under our configurations, the optimum appears near $E_0 \approx 40~\text{mV/m}$ and $B_\text{bias} \approx 0.3~\text{mT}$, yielding a quantum-limited angle resolution of approximately $0.25^\circ$. (iii) \textit{Strong-field region:} Further increasing either $E_0$ or $B_\text{bias}$ separates the Zeeman shifts beyond the EIT linewidth, preventing the accumulation of geometric information and hence degrading the QCRB again.

Fig.~\ref{QCRB_vs_E_and_B_3d} visualizes the QCRB surface across ${E_0} \in \left[ {1~\text{mV/m},1~\text{V/m}} \right]$ and $B_\text{bias} \in \left[ 10~\mu \text{T}, 10~\text{mT} \right] $. As it transpires, the 3D view exhibits a narrow ``valley" of minimum QCRB, where the receiver operates at its quantum optimum. Decreasing either $E_0$ or $B_\text{bias}$ leads to under-splitting of the EIT peaks and thus yields poor sensitivity; increasing either $E_0$ or $B_\text{bias}$ over-splits the resonance and again degrades the angle resolution. Crucially, this valley spans exactly those specific $\left( E_0-B_\text{bias}\right) $ combinations that preserve the EIT resonance linewidth, so that one can tune the fields to the optimum without broadening the EIT lineshape and hence without degrading the sensor's frequency-response. This 3D visualization thus provides a practical guide: stay within the valley to maintain sub-$0.2^\circ$ angle resolution, while retaining the full spectral agility of the Rydberg atomic receiver.

\subsection{Performance Comparison}

\begin{figure}[t]
	\centering
	\includegraphics[width=0.476\textwidth]{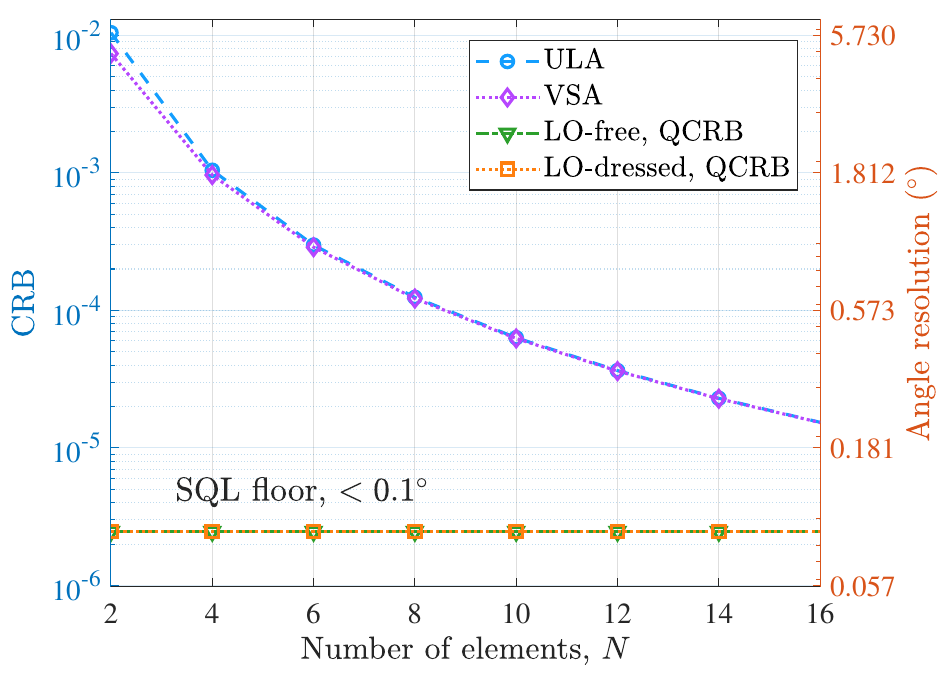}
	\caption{CRB performance versus the number of elements.} \label{CRB_vs_N}
\end{figure}

In this subsection, we juxtapose the fundamental DoA-detection limits of a Rydberg atomic receiver having a single vapor cell with two well-known RF front-end architectures: (i) a conventional half-wavelength ULA-based scalar array, and (ii) an electromagnetic vector-sensor array (VSA)~\cite{VSA}. 
\begin{itemize}
\item ULA: For an $N$-element ULA, the CRB on the one-dimensional DoA estimate is given by~\cite{ULA-CRB1}
\begin{equation}\label{CRB_ULA}
	{\text{CR}}{{\text{B}}_{{\text{ULA}}}} \ge {\left( {{\text{2}} \cdot {\text{SNR}} \cdot {k^2}{{\mathfrak{d}}^2}{{\cos }^2}\theta \sum\nolimits_{n = 0}^{N - 1} {{n^2}} } \right)^{ - 1}},
\end{equation}
where $k=2 \pi / \lambda$ denotes the wavenumber and ${\mathfrak{d}} = \lambda / 2$ is the inter-element spacing. The single-element SNR in (\ref{CRB_ULA}) can be expressed as $\text{SNR} = {{{P_{{\text{Rx}}}}}} / {{\sigma _{{\text{AWGN}}}^2}}$, where ${{P_{{\text{Rx}}}}}$ is the received power and $\sigma _{{\text{AWGN}}}^2$ denotes the power of the additive white-Gaussian-noise (AWGN).

\item VSA: The VSA being considered is a co-linear array of the same $N$ physical positions, each position housing a tri-axial electric-dipole triplet that simultaneously records the three orthogonal electric-field components. Under the identical SNR and spacing configurations, the corresponding CRB becomes~\cite{VSA}
\begin{equation}\label{CRB_VSA}
	{\text{CR}}{{\text{B}}_{{\text{VSA}}}} \ge {\left[ {{\text{2}} \cdot {\text{SNR}} \cdot \Re \left\{ {{{\left( {\frac{{\partial {\mathbf{a}}}}{{\partial \theta }}} \right)}^{H}}\Xi \left( {\frac{{\partial {\mathbf{a}}}}{{\partial \theta }}} \right)} \right\}} \right]^{ - 1}},
\end{equation}
where $\Xi  = {\mathbf{I}} - {\mathbf{a}}\left( \theta  \right){\left[ {{{\mathbf{a}}^H}\left( \theta  \right){\mathbf{a}}\left( \theta  \right)} \right]^{ - 1}}{{\mathbf{a}}^H}\left( \theta  \right)$ and ${\mathbf{a}}\left( \theta  \right)$ represents the array response.
\end{itemize}
As regards to the Rydberg atomic receiver, we consider two architectures: LO-free and LO-dressed (also known as heterodyne~\cite{chen-arxiv,QS-16}) regimes. Their structural difference lies in the presence (LO‑dressed) or absence (LO‑free) of the local oscillator (LO) drive, as detailed in our earlier work~\cite{chen-arxiv}. The DoA is obtained by ${\theta ^{{\text{DoA}}}} = {\text{atan2}}\left( {{{\left( {{\mathbf{E}} \times {{\mathbf{B}}_{{\text{RF}}}}} \right)}_y},{{\left( {{\mathbf{E}} \times {{\mathbf{B}}_{{\text{RF}}}}} \right)}_x}} \right)$, which is a smooth function. The two-parameter QFIM, obtained from the E1 and M1 transition spectra, is hence convert to a scalar bound via error propagation
\begin{align}
{\text{CRB}}\left( {{\theta ^{{\text{DoA}}}}} \right) \ge {\mathbf{J}}{{\mathbf{F}}^{ - 1}}{{\mathbf{J}}^T},
\end{align}
where ${\mathbf{J}} = \left[ {\begin{array}{*{20}{c}}
		{{\partial _{{\theta _{{\text{RF}}}}}}{\theta ^{{\text{DoA}}}}}&{{\partial _{{\theta _B}}}{\theta ^{{\text{DoA}}}}} 
\end{array}} \right]$ is the Jacobi matrix. The CRB is normalized to the same integration time and bandwidth as the array-based case, allowing a direct comparison between the Rydberg QCRB and the array-based CRB under identical conditions.

\begin{figure}[t]
	\centering
	\includegraphics[width=0.49\textwidth]{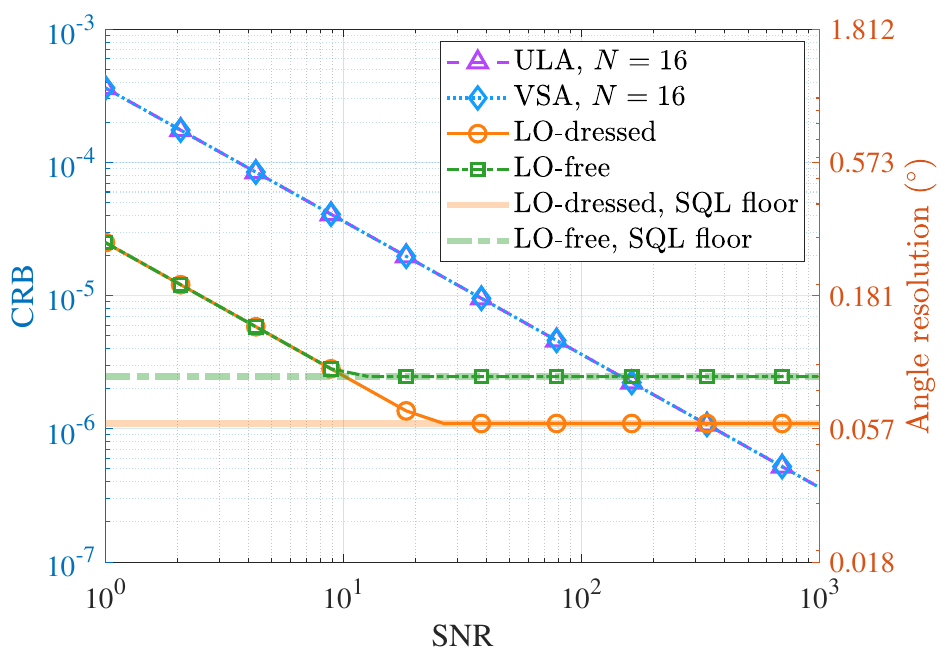}
	\caption{CRB versus SNR in terms of the Rydberg atomic receiver and the ULA-based RF receiver.} \label{CRB_vs_SNR}
\end{figure}

As it transpires from Fig.~\ref{CRB_vs_N}, the theoretical CRB achieved by the ULA- and VSA-based RF receiver scales as $N^{-3}$, exhibiting a steep trend. More explicitly, each doubling of the elements reduces the variance by roughly an order of magnitude. 
For the VSA, the availability of three orthogonal field components per position adds a polarization-derived term to the Fisher information. However, this additional term contributes only a small constant-factor improvement after projection onto the subspace orthogonal to the steering vector. Consequently, the VSA curve is numerically almost indistinguishable from the ULA curve on the plotted scale, and the slope remains determined by the aperture-limited $N^{-3}$ law rather than by the sensor vector dimensionality.
In contrast, regarding the Rydberg atomic receiver, its fundamental limit is determined by the slope of the EIT response and therefore by the intrinsic optical and atomic parameters, such as the dipole moment $\hat{d}$, coherence time $T_2$, dephasing rates $\gamma_{ij}$, atom number $N_\text{atoms}$, and other related quantities, rather than by its physical size. Fig.~\ref{CRB_vs_N} demonstrates that a single vapor cell already achieves sub-degree resolution, outperforming both the classical ULA and VSA up to at least $N \le 16$, while altogether avoiding the hardware scaling and calibration complexity owing to the array aperture growth.


Fig.~\ref{CRB_vs_SNR} compares the CRB of array-based receivers and Rydberg atomic receivers as functions of SNR. As expected, the array-based (ULA and VSA) bounds decrease approximately as $\mathrm{SNR}^{-1}$ under the ideal AWGN model adopted here. 
By contrast, the Rydberg receiver exhibits two regimes. At low to moderate SNR (order $10$), both LO-free and LO-dressed tracks decrease with SNR and outperform the ULA/VSA because their information is set by the EIT dispersion slope rather than by array aperture. Once the SNR enters the tens-to-few-tens range, the QCRB tends to a floor and asymptotically approaches the SQL. Beyond this point, improving the SNR cannot translate into finer angle resolution, because the quantum Fisher information is already saturated. Further gains would require increasing the number of atoms, shortening the measurement cycle or enhancing the optical detection efficiency. Additionally, at low SNR, the LO-dressed and LO-free architectures exhibit similar performance. However, owing to its enhanced sensitivity~\cite{chen-arxiv,QS-34,QS-16}, the LO-dressed scheme features an improved SQL, indicating greater SNR potential and thus a tighter QCRB performance.


\begin{figure}[t]
	\centering
	\includegraphics[width=0.49\textwidth]{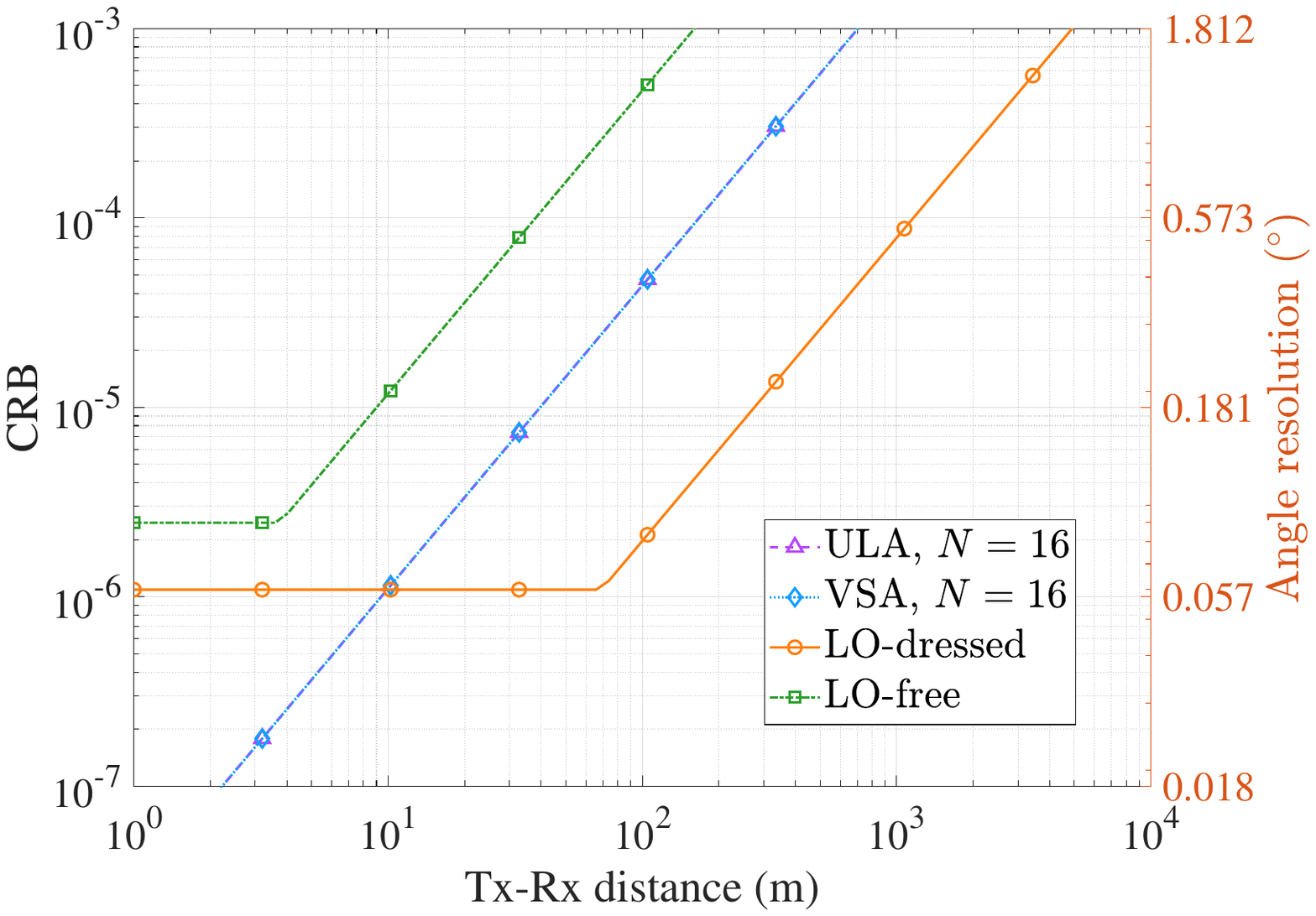}
	\caption{CRB performance versus the Tx-Rx distance.} \label{CRB_vs_dist}
\end{figure}

\begin{figure}[t]
	\centering
	\includegraphics[width=0.49\textwidth]{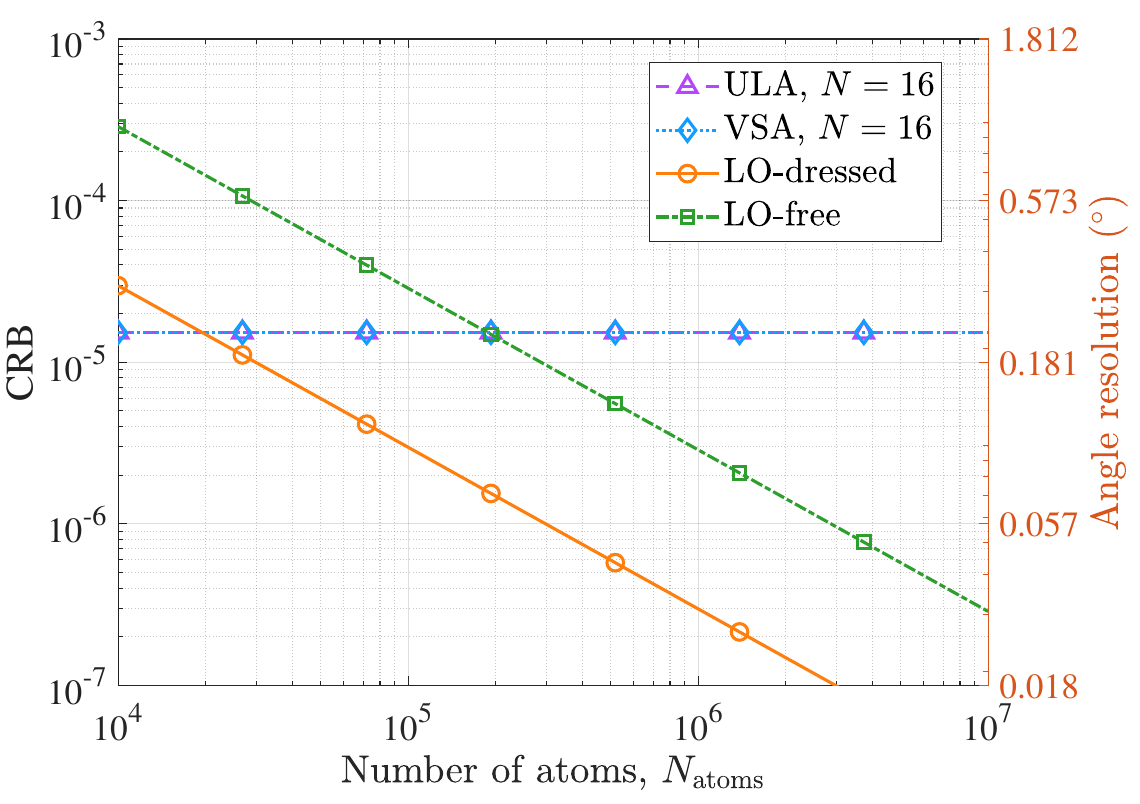}
	\caption{CRB performance versus the number of atoms.} \label{CRB_vs_Natoms}
\end{figure}

Fig.~\ref{CRB_vs_dist} plots the CRB performance versus the Tx–Rx distance for different receiver regimes. As observed,
the LO-dressed Rydberg atomic receiver achieves the best CRB performance, particularly at larger Tx-Rx distances. This is because the transmission distance primarily affects the received power. Given the same transmission distance, the received signals experience the same fading channel before arriving at the receiver front-end. However, receivers differ in their SNR response to identical received power, with the LO-dressed configuration exhibiting the best SNR response. One may refer to our earlier work for detailed analysis~\cite{chen-arxiv}. Accordingly, enhanced SNR leads to improved CRB performance, provided the SQL floor has not yet been reached.
Furthermore, both Rydberg atomic receivers clamp to their SQL-determined floors at short ranges, so further SNR increases do not improve the bound. In this saturation regime, the array-based solutions, i.e., ULA and VSA, outperform the Rydberg atomic receivers.

Fig.~\ref{CRB_vs_Natoms} shows the CRB performance varying with the number of atoms $N_{\text{atoms}}$.
In the case of Rydberg atomic receivers, the QCRB decreases as $N_{\text{atoms}}$ increases, which is consistent with an SQL imposed by intrinsic optical-atomic parameters, i.e., number of atoms $N_\text{atoms}$, rather than by its geometry aperture. Accordingly, one the quantum sensing system nears the SQL floor, further increasing in the SNR may lead to marginal improvements. Enhancements under high-SNR conditions should therefore focus on optimizing intrinsic parameters and optical detection efficiency rather than relying solely on increased RF power.

\section{Conclusions}\label{Sec_VII}

We have conceived a scheme for identifying the DoA of an incident RF signal with sub-0.1$^{\circ}$ precision by relying on a single Rydberg atomic vapor cell. The core idea is that a plane electromagnetic wave is fully characterized by its orthogonal electric- and magnetic-field vectors. Based on this, the Zeeman-resolved EIT spectra can be decoded for extracting both the electric-field polarization angle and RF magnetic-field orientation. An analytic expression of this pair of signatures directly reconstructs the incident wave vector. Furthermore, we have demonstrated that the QFIM places the Rydberg counterpart on the same theoretical footing as classical array-based, such as ULA and VSA, RF receivers and specifies a QCRB that depends only on an information-centric SNR. Our simulation results reveal that, in terms of the detection precision quantified by  CRB, the Rydberg atomic receiver outperforms 16-element array-based receivers across the low-SNR regime and remains robust to performance erosion even though the RF transition is over-driven.

Future efforts will be devoted to integrated photonic-atomic designs to miniaturize the associated vapor cell architectures, and to extend the dynamic range for high-power RF environments. Experimental validation in multipath-rich scenarios and hybrid quantum-classical fusion architectures also present promising avenues. By bridging quantum sensing and wireless communications, this research opens a transformative path toward atomically precise radio, where the EM field control operates at the fundamental limits of quantum mechanics.

\appendices
\section{Proof of Eq.~(\ref{rho_approx})}\label{appendix_proof_rho}

In the weak-probe regime ${\Omega _p} \ll {\gamma _{21}},{\Omega _c}$, the
$\theta$-dependence of the steady-state density matrix is weak, i.e., the
probe-induced population redistribution constitutes only a small perturbation. For an algebraically convenient full-rank reference,
we write the exact decomposition as follows
\begin{subequations}
\begin{align}
	&\bm{\rho} = \frac{{\mathbf{I}}}{4} + \bm{\varepsilon}, \\ &\mathrm{Tr}(\bm{\varepsilon})=0,
	\\ &{\left\| \bm{\varepsilon} \right\|_F} \equiv \varepsilon  \ll 1,
\end{align}
\end{subequations}
and likewise 
\begin{equation}
{\bm{\rho}^*} = \frac{{\mathbf{I}}}{4} + {\bm{\varepsilon}}{^*}.
\end{equation}
Then, we have
\begin{equation}
	{\bm{\rho}^*} \otimes {\mathbf{I}} + {\mathbf{I}} \otimes {\bm{\rho}}
	= \underbrace{\frac{1}{2}{{\mathbf{I}}_{16}}}_{\bm{X}}
	+ \underbrace{\left( { {\bm{\varepsilon}}{^*} \otimes {\mathbf{I}} + {\mathbf{I}} \otimes  {\bm{\varepsilon}}} \right)}_{\bm{Y}} .
\end{equation}
Since $ \left\| \bm{Y}  \right\| _F \le 4\epsilon$ and $\left\| \bm{X}^{-1}  \right\|_2 = 2 $, we have ${\left\| {{\bm{X}^{ - 1}}\bm{Y}} \right\|_F} \le 8 \ll 1$.
Inverting $\bm{X}+\bm{Y}$ via the Neumann series yields that
\begin{equation}
	{\left( {\bm{X} + \bm{Y}} \right)^{ - 1}} = {\bm{X}^{ - 1}}\left( {{\mathbf{I}} + \left( { - {\bm{X}^{ - 1}}\bm{Y}} \right) + {{\left( { - {\bm{X}^{ - 1}}\bm{Y}} \right)}^2} + ...} \right).
\end{equation}
Given ${\bm{X}^{ - 1}} = 2{{\mathbf{I}}_{16}}$, the zeroth‐order approximation yields
\begin{equation}
{\left[ {{{\bm{\rho}}^*} \otimes {\mathbf{I}} + {\mathbf{I}} \otimes {\bm{\rho}})} \right]^{ - 1}} \approx 2{{\mathbf{I}}_{16}},
\end{equation}
while the first-order correction is
\begin{equation}
{\left[ {{{\bm{\rho}}^*} \otimes {\mathbf{I}} + {\mathbf{I}} \otimes \bm{\rho}} \right]^{ - 1}} \approx 2{{\mathbf{I}}_{16}} - 4\bm{Y} + \mathcal{O}\left( {{\varepsilon ^2}} \right).
\end{equation}
In the QFIM, the diagonal element obeys ${{\mathbf{F}}_{\theta_i}} \propto {\left[ {{\text{vec}}\left( {{{\mathbf{A}}_{{\theta _i}}}} \right)} \right]^H}{\text{vec}}\left( {{{\mathbf{A}}_{{\theta _i}}}} \right) $, so the first-order term contributes $\mathcal{O}\left( {{\varepsilon}} \right)$. For typical parameters ${\Omega _p}/{\gamma _{21}} \sim {10^{ - 2}}$, one finds $\varepsilon  \lesssim {10^{ - 2}}$ and hence the relative error is below $1\%$. This thus establishes the validity of Eq.~(\ref{rho_approx}).

\bibliographystyle{IEEEtran}
\bibliography{ref/Ref_Rybg_pol}

@Article{terry-trans,
  author  = {Gong, Tierui and Sun, Jiaming and Yuen, Chau and Hu, Guangwei and Zhao, Yufei and Guan, Yong Liang and See, Chong Meng Samson and Debbah, M{\'e}rouane and Hanzo, Lajos},
  journal = {arXiv preprint: 2412.05554},
  title   = {Rydberg Atomic Quantum Receivers for Classical Wireless Communications and Sensing: Their Models and Performance},
  year    = {2024},
  url     = {https://arxiv.org/abs/2412.05554},
}

@Article{QuTip,
  author    = {Johansson, J Robert and Nation, Paul D and Nori, Franco},
  journal   = {Comp. Phys. Comm.},
  title     = {{QuTiP: An open-source Python framework for the dynamics of open quantum systems}},
  year      = {2012},
  month     = apr,
  number    = {8},
  pages     = {1760--1772},
  volume    = {183},
  publisher = {Elsevier},
}

@Article{QS-62,
  author    = {Rath, Aniket and Branciard, Cyril and Minguzzi, Anna and Vermersch, Beno\^{\i}t},
  journal   = {Phys. Rev. Lett.},
  title     = {Quantum {Fisher} Information from Randomized Measurements},
  year      = {2021},
  month     = dec,
  pages     = {260501},
  volume    = {127},
  doi       = {10.1103/PhysRevLett.127.260501},
  issue     = {26},
  numpages  = {6},
  publisher = {American Physical Society},
}

@Article{QS-61,
  author    = {\ifmmode \check{S}\else \v{S}\fi{}afr\'anek, Dominik},
  journal   = {Phys. Rev. A},
  title     = {Simple expression for the quantum {Fisher} information matrix},
  year      = {2018},
  month     = apr,
  pages     = {042322},
  volume    = {97},
  doi       = {10.1103/PhysRevA.97.042322},
  issue     = {4},
  numpages  = {6},
  publisher = {American Physical Society},
}

@Article{QS-55-2003,
  author    = {Ta\ifmmode \breve{\imath}\else \u{\i}\fi{}chenachev, A. V. and Yudin, V. I. and Wynands, R. and St\"ahler, M. and Kitching, J. and Hollberg, L.},
  journal   = {Phys. Rev. A},
  title     = {Theory of dark resonances for alkali-metal vapors in a buffer-gas cell},
  year      = {2003},
  month     = {Mar},
  pages     = {033810},
  volume    = {67},
  doi       = {10.1103/PhysRevA.67.033810},
  issue     = {3},
  numpages  = {11},
  publisher = {American Physical Society},
}

@Article{QS-54,
  author   = {Chopinaud, A. and Pritchard, J.D.},
  journal  = {Phys. Rev. Appl.},
  title    = {Optimal State Choice for {Rydberg}-Atom Microwave Sensors},
  year     = {2021},
  month    = aug,
  pages    = {024008},
  volume   = {16},
  doi      = {10.1103/PhysRevApplied.16.024008},
  issue    = {2},
  numpages = {10},
}

@Article{QS-53,
  author   = {Cox, Kevin and Yudin, Valery I. and Taichenachev, Alexey V. and Novikova, Irina and Mikhailov, Eugeniy E.},
  journal  = {Phys. Rev. A},
  title    = {Measurements of the magnetic field vector using multiple electromagnetically induced transparency resonances in {Rb} vapor},
  year     = {2011},
  month    = jan,
  pages    = {015801},
  volume   = {83},
  doi      = {10.1103/PhysRevA.83.015801},
  issue    = {1},
  numpages = {4},
}

@Article{QS-52,
  author   = {Yudin, V. I. and Taichenachev, A. V. and Dudin, Y. O. and Velichansky, V. L. and Zibrov, A. S. and Zibrov, S. A.},
  journal  = {Phys. Rev. A},
  title    = {Vector magnetometry based on electromagnetically induced transparency in linearly polarized light},
  year     = {2010},
  month    = sep,
  pages    = {033807},
  volume   = {82},
  doi      = {10.1103/PhysRevA.82.033807},
  issue    = {3},
  numpages = {7},
}

@Article{QS-51,
  author    = {Cloutman, Matthew and Chilcott, Matthew and Elliott, Alexander and Otto, J. Susanne and Deb, Amita B. and Kj\ae{}rgaard, Niels},
  journal   = {Phys. Rev. Appl.},
  title     = {{Polarization-insensitive microwave electrometry using Rydberg atoms}},
  year      = {2024},
  month     = apr,
  pages     = {044025},
  volume    = {21},
  doi       = {10.1103/PhysRevApplied.21.044025},
  issue     = {4},
  numpages  = {6},
  publisher = {American Physical Society},
}

@Article{QS-34,
  author    = {Tu, Hai-Tao and Liao, Kai-Yu and Wang, Hong-Lei and Zhu, Yi-Fei and Qiu, Si-Yuan and Jiang, Hao and Huang, Wei and Bian, Wu and Yan, Hui and Zhu, Shi-Liang},
  journal   = {Sci. Adv.},
  title     = {{Approaching the standard quantum limit of a Rydberg-atom microwave electrometer}},
  year      = {2024},
  month     = dec,
  number    = {51},
  pages     = {eads0683},
  volume    = {10},
  doi       = {DOI: 10.1126/sciadv.ads0683},
  publisher = {American Association for the Advancement of Science},
}

@Article{QS-31,
  author    = {Degen, C. L. and Reinhard, F. and Cappellaro, P.},
  journal   = {Rev. Mod. Phys.},
  title     = {Quantum sensing},
  year      = {2017},
  month     = jul,
  pages     = {035002},
  volume    = {89},
  doi       = {10.1103/RevModPhys.89.035002},
  issue     = {3},
  numpages  = {39},
  publisher = {American Physical Society},
}

@Article{QS-3,
  author   = {Fancher, Charles T. and Scherer, David R. and John, Marc C. St. and Marlow, Bonnie L. Schmittberger},
  journal  = {IEEE Trans. Quantum Eng.},
  title    = {Rydberg Atom Electric Field Sensors for Communications and Sensing},
  year     = {2021},
  month    = mar,
  pages    = {1-13},
  volume   = {2},
  doi      = {10.1109/TQE.2021.3065227},
}

@Article{QS-28,
  author    = {Wu, Fengchuan and An, Qiang and Sun, Zhanshan and Fu, Yunqi},
  journal   = {Phys. Rev. A},
  title     = {Linear dynamic range of a {Rydberg-atom} microwave superheterodyne receiver},
  year      = {2023},
  month     = apr,
  pages     = {043108},
  volume    = {107},
  doi       = {10.1103/PhysRevA.107.043108},
  issue     = {4},
  numpages  = {9},
  publisher = {American Physical Society},
}

@Article{QS-27,
  author  = {Simons, Matthew T. and Haddab, Abdulaziz H. and Gordon, Joshua A. and Holloway, Christopher L.},
  journal = {Appl. Phys. Lett.},
  title   = {{A Rydberg atom-based mixer: Measuring the phase of a radio frequency wave}},
  year    = {2019},
  month   = mar,
  number  = {11},
  pages   = {114101},
  volume  = {114},
  doi     = {10.1063/1.5088821},
}

@Article{QS-26,
  author    = {Yuhan Wang and Fengdong Jia and Jianhai Hao and Yue Cui and Fei Zhou and Xiubin Liu and Jiong Mei and Yonghong Yu and Ya Liu and Jian Zhang and Feng Xie and Zhiping Zhong},
  journal   = {Opt. Express},
  title     = {{Precise measurement of microwave polarization using a Rydberg atom-based mixer}},
  year      = {2023},
  month     = mar,
  number    = {6},
  pages     = {10449--10457},
  volume    = {31},
  doi       = {10.1364/OE.485662},
  publisher = {Optica Publishing Group},
}

@Article{QS-24,
  author    = {Sedlacek, J. A. and Schwettmann, A. and K\"ubler, H. and Shaffer, J. P.},
  journal   = {Phys. Rev. Lett.},
  title     = {Atom-Based Vector Microwave Electrometry Using {Rubidium} {Rydberg} Atoms in a Vapor Cell},
  year      = {2013},
  month     = {Aug},
  pages     = {063001},
  volume    = {111},
  doi       = {10.1103/PhysRevLett.111.063001},
  issue     = {6},
  numpages  = {5},
  publisher = {American Physical Society},
}

@Article{QS-22,
  author    = {Schlossberger, Noah and Prajapati, Nikunjkumar and Berweger, Samuel and Rotunno, Andrew P and Artusio-Glimpse, Alexandra B and Simons, Matthew T and Sheikh, Abrar A and Norrgard, Eric B and Eckel, Stephen P and Holloway, Christopher L},
  journal   = {Nat. Rev. Phys.},
  title     = {Rydberg states of alkali atoms in atomic vapour as {SI}-traceable field probes and communications receivers},
  year      = {2024},
  month     = sep,
  doi       = {https://doi.org/10.1038/s42254-024-00756-7},
  publisher = {Nature Publishing Group UK London},
}

@Article{QS-20,
  author    = {Liu, Zong-Kai and Zhang, Li-Hua and Liu, Bang and Zhang, Zheng-Yuan and Guo, Guang-Can and Ding, Dong-Sheng and Shi, Bao-Sen},
  journal   = {Nat. Commun.},
  title     = {Deep learning enhanced {Rydberg} multifrequency microwave recognition},
  year      = {2022},
  month     = apr,
  number    = {1},
  pages     = {1997},
  volume    = {13},
  publisher = {Nature Publishing Group UK London},
}

@Article{QS-16,
  author    = {Jing, Mingyong and Hu, Ying and Ma, Jie and Zhang, Hao and Zhang, Linjie and Xiao, Liantuan and Jia, Suotang},
  journal   = {Nat. Phys.},
  title     = {Atomic superheterodyne receiver based on microwave-dressed {Rydberg} spectroscopy},
  year      = {2020},
  month     = jun,
  number    = {9},
  pages     = {911--915},
  volume    = {16},
  doi       = {https://doi.org/10.1038/s41567-020-0918-5},
  publisher = {Nature Publishing Group UK London},
}

@Article{QS-15,
  author  = {Rotunno, Andrew P. and Holloway, Christopher L. and Prajapati, Nikunjkumar and Berweger, Samuel and Artusio-Glimpse, Alexandra B. and Brown, Roger and Simons, Matthew and Robinson, Amy K. and Kayim, Baran N. and Viray, Michael A. and Jones, Jasmine F. and Sawyer, Brian C. and Wyllie, Robert and Walker, Thad and Ziolkowski, Richard W. and Jefferts, Steven R. and Geibel, Steven and Wheeler, Jonathan and Imhof, Eric},
  journal = {J. Appl. Phys.},
  title   = {{Investigating electromagnetically induced transparency spectral lineshape distortion due to non-uniform fields in Rydberg-atom electrometry}},
  year    = {2023},
  issn    = {0021-8979},
  month   = aug,
  number  = {8},
  pages   = {084401},
  volume  = {134},
  doi     = {10.1063/5.0161213},
}

@Article{QS-12,
  author    = {Sedlacek, Jonathon A and Schwettmann, Arne and K{\"u}bler, Harald and L{\"o}w, Robert and Pfau, Tilman and Shaffer, James P},
  journal   = {Nat. Phys.},
  title     = {Microwave electrometry with {Rydberg} atoms in a vapour cell using bright atomic resonances},
  year      = {2012},
  month     = nov,
  number    = {11},
  pages     = {819--824},
  volume    = {8},
  publisher = {Nature Publishing Group UK London},
}

@Article{QS-11,
  author  = {Robinson, Amy K. and Prajapati, Nikunjkumar and Senic, Damir and Simons, Matthew T. and Holloway, Christopher L.},
  journal = {Appl. Phys. Lett.},
  title   = {{Determining the angle-of-arrival of a radio-frequency source with a Rydberg atom-based sensor}},
  year    = {2021},
  issn    = {0003-6951},
  month   = mar,
  number  = {11},
  pages   = {114001},
  volume  = {118},
  doi     = {10.1063/5.0045601},
}

@Article{QS-101,
  author   = {Hanzo, Lajos and Babar, Zunaira and Cai, Zhenyu and Chandra, Daryus and Djordjevic, Ivan B. and Koczor, Balint and Ng, Soon Xin and Razavi, Mohsen and Simeone, Osvaldo},
  journal  = {Proc. IEEE},
  title    = {Quantum Information Processing, Sensing, and Communications: Their Myths, Realities, and Futures},
  year     = {to appear, 2025},
  doi      = {10.1109/JPROC.2024.3510394},
}

@Article{QS-1,
  author   = {Cui, Mingyao and Zeng, Qunsong and Huang, Kaibin},
  journal  = {IEEE J. Sel. Areas Commun.},
  title    = {Towards Atomic {MIMO} Receivers},
  year     = {2025},
  month    = mar,
  number   = {3},
  pages    = {659-673},
  volume   = {43},
  doi      = {10.1109/JSAC.2025.3531528},
}

@Misc{3-j,
  author = {A. J. Stone},
  title  = {Wigner $3j$, $6j$, and $9j$ Symbols},
  url    = {https://www-stone.ch.cam.ac.uk/wigner.shtml},
}

@Article{W-E_theorem,
  author  = {Ginibre, Jean},
  journal = {J. Math. Phys.},
  title   = {Wigner‐Eckart Theorem and Simple Lie Groups},
  year    = {1963},
  issn    = {0022-2488},
  month   = may,
  number  = {5},
  pages   = {720-726},
  volume  = {4},
  doi     = {10.1063/1.1704010},
}

@Article{QS-56,
  author  = {Talashila, Rajavardhan and Watterson, William J and Moser, Benjamin L and Gordon, Joshua A and Artusio-Glimpse, Alexandra B and Prajapati, Nikunjkumar and Schlossberger, Noah and Simons, Matthew T and Holloway, Christopher L},
  journal = {arXiv preprint: 2502.09835},
  title   = {Determining angle of arrival of radio frequency fields using subwavelength, amplitude-only measurements of standing waves in a {Rydberg atom} sensor},
  year    = {2025},
  url     = {https://arxiv.org/abs/2502.09835},
}

@Article{C-G_coefficients,
  author   = {McNamee, P. and J., S. and Chilton, Frank},
  journal  = {Rev. Mod. Phys.},
  title    = {Tables of {Clebsch-Gordan} Coefficients of S${\mathrm{U}}_{3}$},
  year     = {1964},
  month    = sep,
  pages    = {1005--1024},
  volume   = {36},
  doi      = {10.1103/RevModPhys.36.1005},
  issue    = {4},
  numpages = {0},
}

@Article{Zeeman_shift,
  author    = {Rinaldi, R. and Giugno, P. V. and Cingolani, R. and Lipsanen, H. and Sopanen, M. and Tulkki, J. and Ahopelto, J.},
  journal   = {Phys. Rev. Lett.},
  title     = {Zeeman Effect in Parabolic Quantum Dots},
  year      = {1996},
  month     = {Jul},
  pages     = {342--345},
  volume    = {77},
  doi       = {10.1103/PhysRevLett.77.342},
  issue     = {2},
  numpages  = {0},
  publisher = {American Physical Society},
}

@Article{chen-arxiv,
  author  = {Chen, Yuanbin and Guo, Xufeng and Yuen, Chau and Zhao, Yufei and Guan, Yong Liang and See, Chong Meng Samson and D{\'e}bbah, Merouane and Hanzo, Lajos},
  journal = {arXiv preprint: 2501.11842},
  title   = {Harnessing {Rydberg} Atomic Receivers: From Quantum Physics to Wireless Communications},
  year    = {2025},
  url     = {https://arxiv.org/abs/2501.11842},
}

@Article{2004_science,
  author    = {Giovannetti, Vittorio and Lloyd, Seth and Maccone, Lorenzo},
  journal   = {Science},
  title     = {Quantum-enhanced measurements: beating the standard quantum limit},
  year      = {2004},
  month     = nov,
  number    = {5700},
  pages     = {1330--1336},
  volume    = {306},
  publisher = {American Association for the Advancement of Science},
}

@Article{QS-32,
  author  = {Santamaria-Botello, Gabriel and Verploegh, Shane and Bottomley, Eric and Popovic, Zoya},
  journal = {arXiv preprint: 2209.00908},
  title   = {Comparison of noise temperature of {Rydberg-atom} and electronic microwave receivers},
  year    = {2022},
  url     = {https://arxiv.org/abs/2209.00908},
}

@Article{ULA-CRB1,
  author   = {Gazzah, H. and Marcos, S.},
  journal  = {IEEE Trans. Signal Process.},
  title    = {{Cramer-Rao} bounds for antenna array design},
  year     = {2006},
  month    = jan,
  number   = {1},
  pages    = {336-345},
  volume   = {54},
  doi      = {10.1109/TSP.2005.861091},
}

@Article{terry-wcl,
  author  = {Gong, Tierui and Yuen, Chau and See, Chong Meng Samson and Debbah, M{\'e}rouane and Hanzo, Lajos},
  journal = {arXiv preprint: 2501.02820},
  title   = {Rydberg Atomic Quantum Receivers for Multi-Target {DOA} Estimation},
  year    = {2025},
  url     = {https://arxiv.org/abs/2501.02820},
}

@Article{terry-WCM,
  author   = {Gong, Tierui and Chandra, Aveek and Yuen, Chau and Guan, Yong Liang and Dumke, Rainer and See, Chong Meng Samson and Debbah, Merouane and Hanzo, Lajos},
  journal  = {IEEE Wirel. Commun.},
  title    = {Rydberg Atomic Quantum Receivers for Classical Wireless Communication and Sensing},
  year     = {to appear, 2025},
  doi      = {10.1109/MWC.015.2400381},
}

@Article{QS-9-TAP-2024,
  author   = {Mao, Ruiqi and Lin, Yi and Zhou, Aojie and Yang, Kai and Fu, Yunqi},
  journal  = {IEEE Trans. Antennas Propag.},
  title    = {Shortwave Ultrahigh-Sensitivity {Rydberg} Atomic Electric Field Sensing Based on a Subminiature Resonator},
  year     = {2024},
  month    = nov,
  number   = {11},
  pages    = {8165-8172},
  volume   = {72},
  doi      = {10.1109/TAP.2024.3439862},
}

@Article{QS-41,
  author   = {Kim, Hanvit and Park, Hyunwoo and Kim, Sunwoo},
  journal  = {IEEE Wireless Commun. Lett.},
  title    = {Quantum-{MUSIC}: Multiple Signal Classification for Quantum Wireless Sensing},
  year     = {to appear, 2025},
  doi      = {10.1109/LWC.2025.3549767},
}

@Article{QS-64,
  author    = {Beloy, K.},
  journal   = {Phys. Rev. Lett.},
  title     = {Trap-Induced {AC} {Zeeman} Shift of the {Thorium}-229 Nuclear Clock Frequency},
  year      = {2023},
  month     = {Mar},
  pages     = {103201},
  volume    = {130},
  doi       = {10.1103/PhysRevLett.130.103201},
  issue     = {10},
  numpages  = {6},
  publisher = {American Physical Society},
}

@Article{QS-9-TAP-2025,
  author   = {Wu, Bo and Mao, Ruiqi and Sang, Di and Sun, Zhanshan and Liu, Yi and Lin, Yi and An, Qiang and Fu, Yunqi},
  journal  = {IEEE Trans. Antennas Propag.},
  title    = {Enhancing Sensitivity of Atomic Microwave Receivers Based on Optimal Laser Arrays},
  year     = {2025},
  month    = feb,
  number   = {2},
  pages    = {793-806},
  volume   = {73},
  doi      = {10.1109/TAP.2024.3486553},
}

@Article{QS-30,
  author  = {Bor{\'o}wka, Sebastian and Pylypenko, Uliana and Mazelanik, Mateusz and Parniak, Micha{\l}},
  journal = {Nat. Photonics},
  title   = {Continuous wideband microwave-to-optical converter based on room-temperature {Rydberg} atoms},
  year    = {2024},
  month   = oct,
  number  = {1},
  pages   = {32--38},
  volume  = {18},
}

@Article{VSA,
  author   = {Nehorai, A. and Paldi, E.},
  journal  = {IEEE Trans. Signal Process.},
  title    = {Vector-sensor array processing for electromagnetic source localization},
  year     = {1994},
  month    = feb,
  number   = {2},
  pages    = {376-398},
  volume   = {42},
  doi      = {10.1109/78.275610},
}

@Article{QS-58,
  author    = {Xinheng Li and Yue Cui and Jianhai Hao and Fei Zhou and Yuxiang Wang and Fengdong Jia and Jian Zhang and Feng Xie and Zhiping Zhong},
  journal   = {Opt. Express},
  title     = {{Magnetic-field-induced splitting of Rydberg Electromagnetically Induced Transparency and Autler-Townes spectra in 87Rb vapor cell}},
  year      = {2023},
  month     = nov,
  number    = {23},
  pages     = {38165--38178},
  volume    = {31},
  doi       = {10.1364/OE.505488},
  publisher = {Optica Publishing Group},
}

@Article{Yan_2023,
  author    = {Yang Yan and Jinpeng Yuan and Linjie Zhang and Liantuan Xiao and Suotang Jia and Lirong Wang},
  journal   = {Opt. Lett.},
  title     = {{Three-dimensional location system based on an L-shaped array of Rydberg atomic receivers}},
  year      = {2023},
  month     = aug,
  number    = {15},
  pages     = {3945--3948},
  volume    = {48},
  doi       = {10.1364/OL.496057},
  publisher = {Optica Publishing Group},
}

@Article{Yuan_2023,
  author    = {Yuan, Jinpeng and Yang, Wenguang and Jing, Mingyong and Zhang, Hao and Jiao, Yuechun and Li, Weibin and Zhang, Linjie and Xiao, Liantuan and Jia, Suotang},
  journal   = {Rep. Prog. Phys.},
  title     = {Quantum sensing of microwave electric fields based on {Rydberg} atoms},
  year      = {2023},
  month     = sep,
  number    = {10},
  pages     = {106001},
  volume    = {86},
  doi       = {10.1088/1361-6633/acf22f},
  publisher = {IOP Publishing},
}

\begin{IEEEbiography}[{\includegraphics[width=1in,height=1.25in,clip,keepaspectratio]{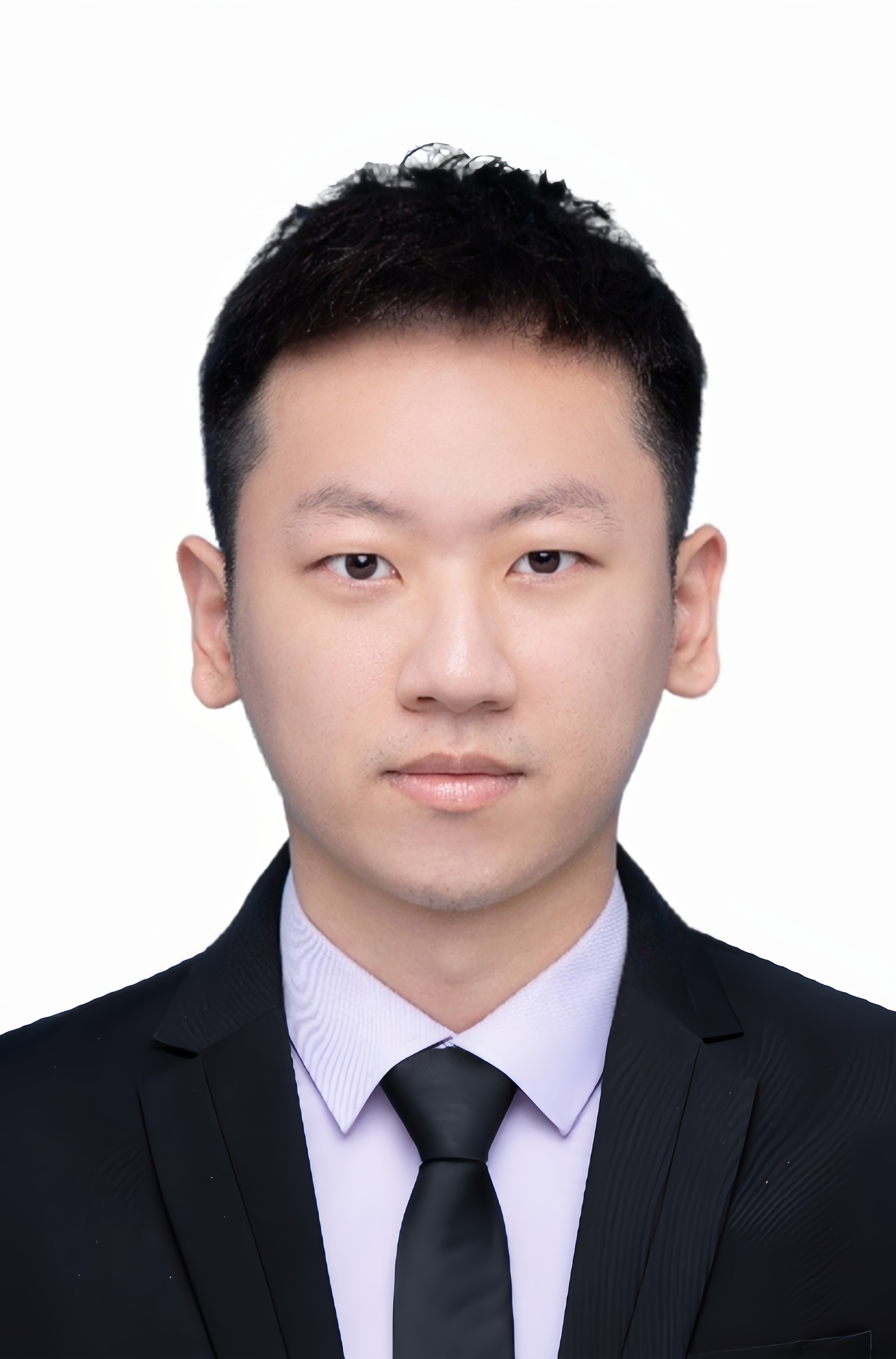}}]{Yuanbin Chen} is a Research Fellow with the School of Electrical and Electronic Engineering, Nanyang Technological University (NTU), Singapore. He received the B.S. degree in Communications Engineering from Beijing Jiaotong University, Beijing, China, in 2019, and the Ph.D. degree in Information and Communication Systems from Beijing University of Posts and Telecommunications (BUPT), Beijing, China, in 2024. He graduated with honors from both institutions.

He was selected in the Stanford/Elsevier Top 2\% Scientists list in 2024 and 2025 and received the Excellent Ph.D. Dissertation Award from the Chinese Institute of Electronics Education Society (2025). He received the 2025 IEEE IWCMC Best Paper Award and the 2023 IEEE GLOBECOM Best Workshop Paper Award. His current research interests include quantum sensing, Rydberg atomic receivers, and 6G advanced MIMO technologies.
\end{IEEEbiography}

\begin{IEEEbiography}[{\includegraphics[width=1in,height=1.25in,clip,keepaspectratio]{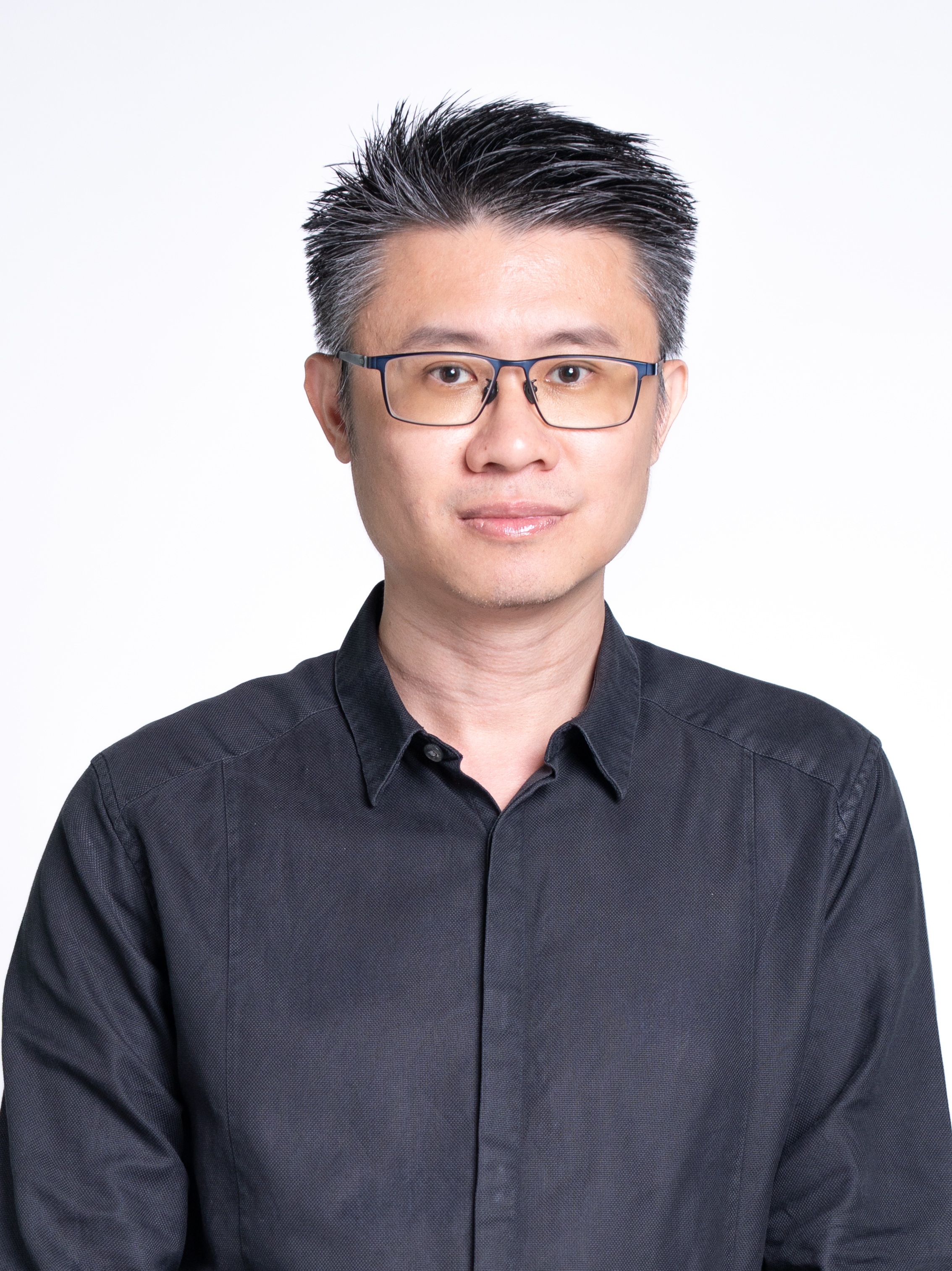}}]{Chau Yuen} received the B.Eng. and Ph.D. degrees from Nanyang Technological University, Singapore, in 2000 and 2004, respectively. He was a Post-Doctoral Fellow with Lucent Technologies Bell Labs, Murray Hill, in 2005. From 2006 to 2010, he was with the Institute for Infocomm Research, Singapore. From 2010 to 2023, he was with the Engineering Product Development Pillar, Singapore University of Technology and Design. Since 2023, he has been with the School of Electrical and Electronic Engineering, Nanyang Technological University, currently he is Provost’s Chair in Wireless Communications, Assistant Dean in Graduate College, and Cluster Director for Sustainable Built Environment at ER@IN. 
	
Dr. Yuen received IEEE Communications Society Leonard G. Abraham Prize (2024), IEEE Communications Society Best Tutorial Paper Award (2024), IEEE Communications Society Fred W. Ellersick Prize (2023), IEEE Marconi Prize Paper Award in Wireless Communications (2021), IEEE APB Outstanding Paper Award (2023), and EURASIP Best Paper Award for JOURNAL ON WIRELESS COMMUNICATIONS AND NETWORKING (2021).
	
Dr. Yuen current serves as an Editor for IEEE TRANSACTIONS ON VEHICULAR TECHNOLOGY, IEEE TRANSACTIONS ON NEURAL NETWORKS AND LEARNING SYSTEMS, and IEEE TRANSACTIONS ON NETWORK SCIENCE AND ENGINEERING, where he was awarded as IEEE TNSE Excellent Editor Award 2025, 2024 and 2022, and Top Associate Editor for TVT from 2009 to 2015.
	
He is listed as Top 2\% Scientists by Stanford University, and also a Highly Cited Researcher by Clarivate Web of Science from 2022.
\end{IEEEbiography}

\begin{IEEEbiography}[{\includegraphics[width=1in,height=1.25in,clip,keepaspectratio]{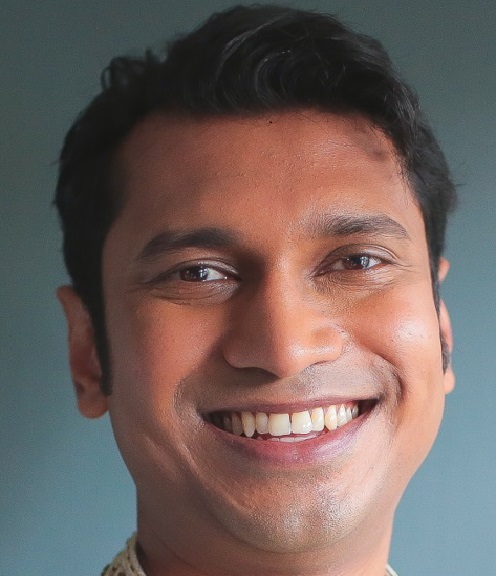}}]{Darmindra Arumugam} is a Principal Research Technologist at NASA's Jet Propulsion Laboratory (JPL), California Institute of Technology. His research focuses on electromagnetics, remote sensing, and quantum sensors, with an emphasis on Rydberg atom–based optical-to-microwave detection for signals-of-opportunity and passive spectrum sensing applications. He has authored more than 75 peer-reviewed publications and holds over 25 patents, contributing to sensing technologies for NASA missions and space-based systems, including applications in spectrum monitoring and GPS-denied navigation. Dr. Arumugam received the Ph.D. degree in electrical and computer engineering, with a focus on applied physics, from Carnegie Mellon University in 2011.
\end{IEEEbiography}

\begin{IEEEbiography}[{\includegraphics[width=1in,height=1.25in,clip,keepaspectratio]{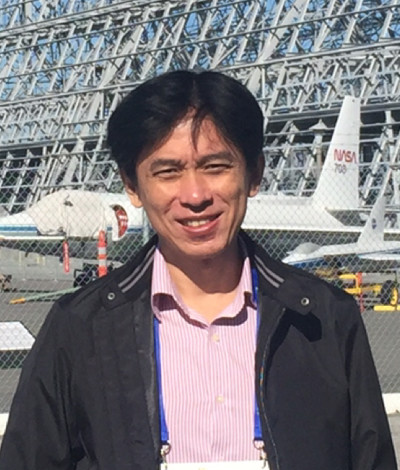}}]{Chong Meng Samson See} received the Diploma in Electronics and Communications Engineering (with Merit) from Singapore Polytechnic, Singapore, in 1988, and the M.Sc. in Digital Communication Systems and the Ph.D. in Electrical Engineering from Loughborough University of Technology, Loughborough, U.K., in 1991 and 1999, respectively.
	
Since 1992, he has been with DSO National Laboratories, Singapore, where he is currently a Distinguished Member of Technical Staff leading the research and development of advanced array signal processing systems and algorithms. He also holds an adjunct appointment as Principal Research Scientist at Temasek Laboratories @ NTU, Singapore. His research interests include statistical signal processing, array signal processing, communications, quantum sensing, and the intersection of signal processing and artificial intelligence.
	
Dr. See holds two issued patents in direction finding. His contributions to defence R\&D have been recognized with the Defence Technology Prize 2015 Individual R\&D Award, the DTP 2014 Team (Engineering) Award, the DTP 2022 Team (Research) Award, and the DSO Kinetic Award in 2011 and 2022. He served as an Associate Editor of the IEEE Transactions on Signal Processing and as a member of the IEEE Sensor Array and Multichannel Signal Processing Technical Committee.
\end{IEEEbiography}

\begin{IEEEbiography}[{\includegraphics[width=1in,height=1.25in,clip,keepaspectratio]{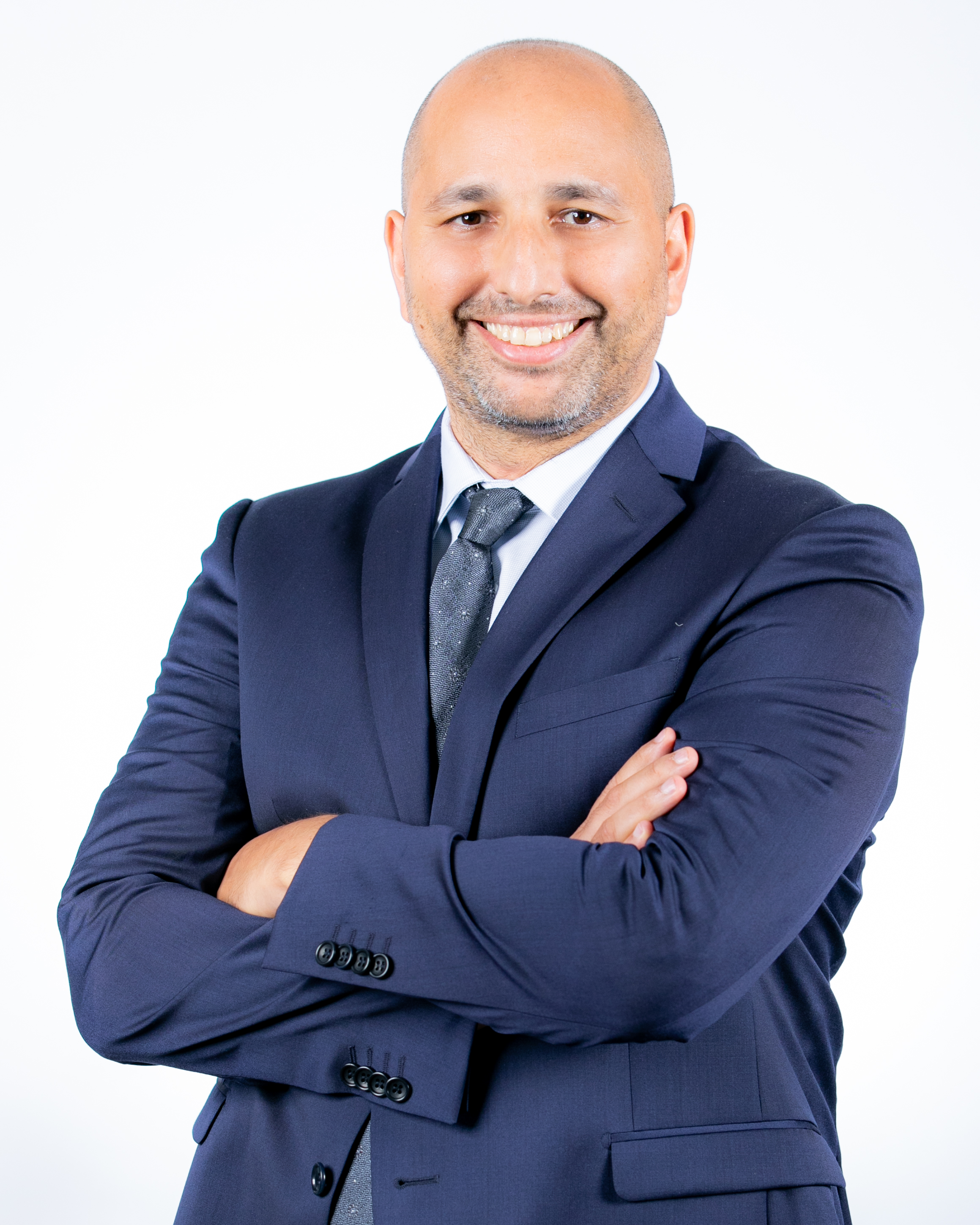}}]{M\'{e}rouane Debbah} is Professor at Khalifa University of Science and Technology in Abu Dhabi and founding Senior Director of KU Digital Future Institute. His research has been lying at the interface of fundamental mathematics, algorithms, statistics, information and communication sciences with a special focus on random matrix theory and learning algorithms. In the Communication field, he has been at the heart of the development of small cells (4G), Massive MIMO (5G) and Large Intelligent Surfaces (6G) technologies. In the AI field, he is known for his work on Large Language Models, distributed AI systems for networks and semantic communications. He received multiple prestigious distinctions, prizes and best paper awards (more than 50 IEEE best paper awards) for his contributions to both fields. He is an IEEE Fellow, a WWRF Fellow, a Eurasip Fellow, an AAIA Fellow, an Institut Louis Bachelier Fellow, an AIIA Fellow  and a Membre émérite SEE. He is actually chair of  the IEEE Large Generative AI Models in Telecom (GenAINet) Emerging Technology Initiative and  a member of the Marconi Prize Selection Advisory Committee.
\end{IEEEbiography}

\begin{IEEEbiography}[{\includegraphics[width=1in,height=1.25in,clip,keepaspectratio]{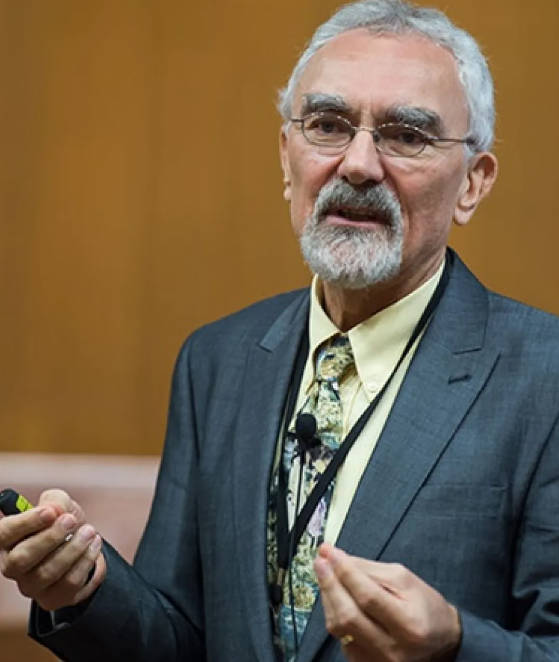}}]{Lajos Hanzo} has co-authored more than 2000 contributions at IEEE Xplore and 19 Wiley-IEEE Press monographs. He is a fellow of the Royal Academy of Engineering, a FIET, a fellow of EURASIP, and a Foreign Member of the Hungarian Academy of Sciences. He was bestowed upon the IEEE Eric Sumner Technical Field Award.
	
\end{IEEEbiography}

\end{document}